\documentclass[12pt,showpacs,floatfix,superscriptaddress,nofootinbib,aps,prc,eqsecnum]{revtex4}
\usepackage[dvips]{graphicx}
\usepackage{latexsym,amssymb,amsmath,epsfig,bm,times,psfrag,subfigure}
\usepackage{color}
\usepackage[bookmarksnumbered,bookmarksopen,colorlinks,citecolor=red,linkcolor=blue]{hyperref}
\renewcommand\a{\alpha}
\renewcommand\b{\beta}
\renewcommand\d{\delta}

\renewcommand\r{\rho}

\renewcommand\t{\tau}

\renewcommand\c{\chi}
\renewcommand\j{\psi}
\renewcommand\o{\omega}
\newcommand\e{\epsilon}
\newcommand\g{\gamma}

\newcommand\m{\mu}
\newcommand\n{\nu}

\newcommand\p{\pi}

\newcommand\s{\sigma}
\newcommand\f{\phi}
\newcommand\w{\eta}
\newcommand\ve{\varepsilon}

\renewcommand\L{\Lambda}

\renewcommand\O{\Omega}

\newcommand\D{\Delta}

\newcommand\F{\Phi}

\newcommand{\fig}[1]{Fig.~\ref{#1}}
\newcommand{\eq}[1]{Eq.~(\ref{#1})}

\newcommand{\sect}[1]{Sec.~\ref{#1}}

\newcommand\lb{\left(}
\newcommand\rb{\right)}
\newcommand\ls{\left[}
\newcommand\rs{\right]}

\newcommand{\lan}{\langle}
\newcommand{\ran}{\rangle}

\newcommand\ra{\rightarrow}

\newcommand{\non}{\nonumber\\}
\newcommand\pt{\partial}

\newcommand{\diag}{{\rm{diag}}}

\newcommand{\bx}{{\bm x}}

\newcommand{\bp}{{\bm p}}
\newcommand{\bk}{{\bm k}}

\newcommand{\bv}{{\bm v}}

\newcommand{\by}{{\bm y}}

\newcommand{\bB}{{\bm B}}
\newcommand{\bE}{{\bm E}}

\renewcommand{\part}{{\rm part}}

\renewcommand{\vec}{\boldsymbol}
\newcommand{\be}{\begin{equation}}
\newcommand{\ee}{\end{equation}}
\newcommand{\bear}{\begin{eqnarray}}
\newcommand{\eear}{\end{eqnarray}}
\newcommand{\ba}{\begin{array}}
\newcommand{\ea}{\end{array}}

\begin{document}

\title{Vorticity in Heavy-Ion Collisions}
\author{Wei-Tian Deng}
\affiliation{School of physics, Huazhong University of Science and Technology, Wuhan 430074, China}
\author{Xu-Guang Huang}
\affiliation{Physics Department and Center for Particle Physics and Field Theory, Fudan University, Shanghai 200433, China.}

\date{\today}

\begin{abstract}
We study the event-by-event generation of flow vorticity in the BNL Relativistic Heavy Ion Collider Au + Au collisions and CERN Large Hadron Collider Pb + Pb collisions by using the HIJING model. Different definitions of the vorticity field and velocity field are considered. A variety of properties of the vorticity are explored, including the impact parameter dependence, the collision energy dependence, the spatial distribution, the event-by-event fluctuation of the magnitude and azimuthal direction, and the time evolution. In addition, the spatial distribution of the flow helicity is also studied.
\end{abstract}
\maketitle

\section {Introduction}\label{intro}
In high-energy heavy-ion collisions, two atomic nuclei collide at relativistic energies such that the energy deposited in the reaction region can be large enough to produce the deconfined quark-gluon matter --- usually called the quark-gluon plasma (QGP). In addition, recent studies revealed that the heavy-ion collisions generate also extremely strong electromagnetic fields. The numerical simulations found that the magnetic fields generated in Au + Au collisions at the top energy currently available at the BNL Relativistic Heavy Ion Collider (RHIC), $\sqrt{s}=200$ GeV, can reach $eB\sim$ several $m_\p^2$ (where $e$ is the absolute value of the electron charge and $m_\p$ is the pion mass) and in the Pb + Pb collisions at the CERN Large Hadron Collider (LHC) energy, $\sqrt{s}=2.76$ TeV, can reach $eB\sim$ several tens of $m_\p^2$~\cite{Skokov:2009qp,Voronyuk:2011jd,Bzdak:2011yy,Deng:2012pc,Mo:2013qya}. The generated electric fields, owing to the event-by-event fluctuation of charge distribution of nucleus or due to the asymmetric collision geometry (e.g., the Cu + Au collisions), can be of the same order of magnitude as the magnetic fields~\cite{Bzdak:2011yy,Deng:2012pc,Hirono:2012rt,Deng:2014uja,Voronyuk:2014rna}. When coupled to the ${\cal P}$ and/or ${\cal C}$ odd domains in the QGP, these strong electromagnetic fields can induce remarkable anomalous transport phenomena, including chiral magnetic effect (CME)~\cite{Kharzeev:2007jp,Fukushima:2008xe}, chiral separation effect (CSE)~\cite{Son:2004tq,Metlitski:2005pr}, chiral magnetic wave (CMW)~\cite{Kharzeev:2010gd,Burnier:2011bf}, chiral electric separation effect (CESE)~\cite{Huang:2013iia,Jiang:2014ura,Pu:2014cwa,Ma:2015isa}, etc. Recently, the measurements performed by STAR Collaboration at RHIC~\cite{Abelev:2009ac,Abelev:2009ad,Wang:2012qs,Adamczyk:2014mzf,Adamczyk:2015eqo} and by ALICE Collaboration at LHC~\cite{Abelev:2012pa,Adam:2015vje} showed features consistent with the expectation of CME and CMW although the experimental observables receive significant contributions from background effects which are still not successfully subtracted; see Refs.~\cite{Kharzeev:2013ffa,Liao:2014ava,Huang:2015oca,Kharzeev:2015znc} for reviews.

The existence of strong magnetic fields suggests that there may be fast rotation and/or large flow vorticity in the produced quark-gluon matter in heavy-ion collisions. In fact, in classical physics, the Larmor's theorem states that the motion of a charged particle in a magnetic field $\bB$ is equivalent to the motion in a rotating frame with angular velocity $\bm \O=-q\bB/(2m)$ (plus an additional centrifugal force) where $m$ and $q$ are the mass and charge of the particle. On the other hand, it is also very natural to expect the appearance of flow vorticity in heavy-ion collisions. Consider a non-central heavy-ion collision with impact parameter $b$. The initial angular momentum $J_0$ of the two nuclei with respect to the collision center is roughly given by $Ab\sqrt{s}/2$ with $A$ the number of nucleons in one nucleus. We can easily estimate the magnitude of $J_0$. For example, for Au + Au collisions at $\sqrt{s}=200$ GeV at $b=10$ fm, $J_0\sim 10^6$; and for Pb + Pb collisions at $\sqrt{s}=2.76$ TeV with
$b=10$ fm, $J_0\sim 10^7$. After the collision, a fraction of the initial angular momentum is retained in the interaction region. This fraction of initial angular momentum manifests itself as a shear of the longitudinal momentum density or velocity field. As a consequence of this shear flow, nonzero local vorticity arises which should be roughly perpendicular to the reaction plane.

Such voticity provides us the possibility to monitor the nontrivial topological sector of quantum chromodynamics (QCD) via the so-called chiral vortical effect (CVE) which is the vortical analog of CME and CSE and represents the generation of vector and axial currents along the vorticity~\cite{Erdmenger:2008rm,Banerjee:2008th,Son:2009tf}. The CVE can be neatly expressed as
\begin{eqnarray}
\label{cve}
{\bm j}&=&\c{\bm\o},\\
\label{cve2}
{\bm j}_5&=&\c_5{\bm\o},
\end{eqnarray}
where ${\bm\o}$ is the flow vorticity, ${\bm j}$ and ${\bm j}_5$ are the vector and chiral currents, respectively. The two conductivities are $\c=N_c\m\m_5/(2\p^2)$ and $\c_5=N_c[T^2/12+(\m^2+\m_5^2)/(4\p^2)]$ with $\m$ the baryon chemical potential, $\m_5$ the chiral chemical potential, and $T$ the temperature. The coupled evolution of the vector and axial currents and densities lead to propagating collective mode called the chiral vortical wave (CVW)~\cite{Jiang:2015cva} which is the vortical analog of CMW. In presence of both vorticity and magnetic field, even complex collective modes, like the chiral heat wave and its mixture with CMW and CVW, can emerge~\cite{Chernodub:2015gxa}. Phenomenologically, the CVE can induce baryon charge separation along the vorticity direction which can be detected via specifically designed two-particle correlation~\cite{Kharzeev:2010gr,Zhao:2014aja} (see also Sec.~\ref{subazi}). The CVW can cause flavor quadrupole in QGP which in turn can lead to elliptic flow splitting
effect for $\L$ baryons that may be experimentally measured~\cite{Jiang:2015cva}. Recently, the STAR Collaboration at RHIC has reported signals that are qualitatively agree with the expectation of the CVE~\cite{Zhao:2014aja}. The flow vorticity may also lead to other novel effects in heavy-ion collisions; see e.g. Refs.~\cite{Liang:2004ph,Gao:2007bc,Betz:2007kg,Huang:2011ru,Rogachevsky:2010ys,Baznat:2013zx,Gao:2012ix,Becattini:2013vja,Csernai:2013vda,Chen:2015hfc,McInnes:2014haa,McInnes:2015kec}.

There were already works that investigated the vorticity in heavy-ion collisions~\cite{Betz:2007kg,Becattini:2007sr,Huang:2011ru,Csernai:2013vda,Csernai:2013bqa,Gao:2014coa,Csernai:2014ywa,Csernai:2014hva,Becattini:2015ska,Teryaev:2015gxa}; some of them will be discussed in the present paper. However, as far as we know, a systematic study of the following issues within a unified framework is still lacking~\footnote{After the main results of the present paper were being completed, we learned that the authors of Ref.~\cite{Jiang:2016woz} also performed detailed numerics to study the vorticity by using AMPT model. Their results have some overlap with ours.}: how large the vorticity can be, how it depends on centrality in different collision systems especially in RHIC Au + Au collisions and in LHC Pb + Pb collisions, how the vorticity is distributed over space and time, how the magnitude and azimuthal direction of the vorticity fluctuate over events, and how its orientation correlates to the matter
geometry. These issues are very important for the understanding of various vorticity-induced effects, e.g., the CVE and CVW in heavy-ion collisions; see the discussions in Sec.~\ref{subazi}. In this paper, we will study these issues in detail in a manner parallel to the previous study of electromagnetic fields in heavy-ion collisions in Refs.~\cite{Deng:2012pc,Bloczynski:2012en,Bloczynski:2013mca,Deng:2014uja,Huang:2015oca}. We will consider different kinds of definition for the vorticity field and velocity field and perform numerical simulations of the generation of vorticity on event-by-event basis by using HIJING model~\cite{Wang:1991hta,Gyulassy:1994ew,Deng:2010mv,Deng:2010xg}. At the mean time, the event-by-event fluctuating participant planes will be also calculated by using HIJING model and the azimuthal correlation between vorticity and participant plane will be studied as well.

This paper is organized as follows. In Sec.~\ref{hydro}, we will give a brief review of some theoretical aspects of the vorticity in hydrodynamics.
In Sec.~\ref{setup}, we will set up our numerical simulation. The numerical results will be presented in Sec.~\ref{globa} and Sec.~\ref{resvor}. A hydrodynamic analysis of the time evolution of the vorticity is presented in Sec.~\ref{sectime}. Finally, we will
summarize the main findings in Sec.~\ref{discu}. Throughout this paper, we use natural units $\hbar=c=k_B=1$ and the metric $g_{\m\n}=g^{\m\n}=\diag(1,-1,-1,-1)$.

\section {Review of vorticity in hydrodynamics}\label{hydro}
\subsection {Non-relativistic case}\label{nonrelat}
In non-relativistic hydrodynamics, the vorticity (pseudo)vector field is defined by~\footnote{This definition follows the convention of classical fluid mechanics, see for example, Ref.~\cite{Landaufluid}. In the literature on chiral vortical effect, a factor $1/2$ is usually inserted in front of the curl to define the vorticity.}
\begin{eqnarray}
\label{defvor1}
\vec\o(\bx,t)=\vec\nabla\times\vec v,
\end{eqnarray}
where $\vec v$ is the flow velocity. Hereafter, we will use $\vec \o_1$ to denote \eq{defvor1} in order to avoid confusion with the vorticity that will be defined in next subsection. The vorticity $\vec \o_1$ is a measure of the local angular velocity of the fluid. For ideal barotropic fluid, i.e., the fluid whose viscosity is negligible and in which the pressure $P$ is a function of the mass density $\r$, i.e., $P=P(\r)$, the evolution of the vorticity is governed by the following vorticity equation,
\begin{eqnarray}
\label{equvor1}
\frac{\pt \vec\o_1}{\pt t}=\vec\nabla\times(\vec v\times\vec\o_1),
\end{eqnarray}
which has the following two remarkable consequences.

(1) The Helmholtz-Kelvin theorem (circulation conservation). This states that the closed contour line integral of the velocity field (called circulation) is conserved as the contour is transported by the flow, i.e.,
\begin{eqnarray}
\frac{d}{dt}\oint\vec v\cdot d \vec x=0,
\end{eqnarray}
where $d/dt$ is understood as the comoving time derivative. Another way to state the Helmholtz-Kelvin theorem is that in an ideal barotropic fluid the vortex lines are comoving with the fluid as if they are frozen in the fluid.

(2) The helicity conservation. From the velocity and vorticity fields, one can construct a pseudoscalar field,
\begin{eqnarray}
\label{nonheli}
h_{\rm f}(\vec x,t)=\vec v\cdot\vec\o_1,
\end{eqnarray}
which is called the helicity density of the flow~\cite{Moffatt:1969}. The integral of $h_{\rm f}$ over the whole space,
\begin{eqnarray}
\label{nonheli2}
{\cal H}_{\rm f}=\int d^3\vec x\; h_{\rm f}=\int d^3\vec x\; \vec v\cdot\vec\o_1,
\end{eqnarray}
is the total helicity of the flow. When the vorticity equation (\ref{equvor1}) holds, the total helicity ${\cal H}_{\rm f}$ is conserved~\cite{Moreau:1961,Moffatt:1969}. Moreover, as first pointed out by Moffatt, the ${\cal H}_{\rm f}$ is actually a topological invariant of the flow --- it measures the degree of intertwist of the vortex lines in the fluid~\cite{Moffatt:1969,Arnold:1998,Moffatt:2014}.

To end this subsection, we note an interesting similarity between the vorticity in an ideal fluid and the magnetic field in a perfectly conducting plasma. Let $\vec B(\vec x,t)$ be the magnetic field and $\vec A(\vec x,t)$ be the vector potential, i.e., $\vec B=\vec\nabla\times\vec A$. The equation that governs the evolution of $\vec B$ in a perfectly conducting plasma reads
\begin{eqnarray}
\label{equmag}
\frac{\pt \vec B}{\pt t}=\vec\nabla\times(\vec v\times\vec B),
\end{eqnarray}
which bears immediate similarity with \eq{equvor1}. In fact, from \eq{equmag} the magnetic frozen-in theorem~\cite{JacksonEM} follows, which states that the magnetic lines are frozen in a perfectly conducting plasma just like that the vortex lines are frozen in an ideal barotropic fluid. Furthermore, by replacing $\vec v$ with $\vec A$ and $\vec\o_1$ with $\vec B$ in \eq{nonheli} and \eq{nonheli2}, one can define the magnetic helicity density
\begin{eqnarray}
h_{\rm M}=\vec A\cdot\vec B
\end{eqnarray}
and the total magnetic helicity in the plasma
\begin{eqnarray}
{\cal H}_{\rm M}=\int d^3\vec x h_{\rm M}=\int d^3\vec x \vec A\cdot\vec B.
\end{eqnarray}
Although $h_{\rm M}$ is gauge dependent, ${\cal H}_{\rm M}$ is a gauge invariant quantity. It is straightforward to show~\cite{Woltjer:1956} that ${\cal H}_{\rm M}$ is a conserved quantity under the time evolution of \eq{equmag} and it is also a topological invariant that measures the degree of intertwist of the magnetic lines~\cite{Arnold:1998,Moffatt:2014}.

\subsection {Relativistic case}\label{relat}
A natural extension of the definition (\ref{defvor1}) to relativistic fluid is
\begin{eqnarray}
\label{defvor2}
\o^\m=\e^{\m\n\r\s}u_\n \pt_\r u_\s=\frac{1}{2}\e^{\m\n\r\s}u_\n \o_{\r\s},
\end{eqnarray}
where $u^\m$ is the 4-velocity of the fluid which is normalized as $u^\m u_\m=1$ with $u_0=\g=1/\sqrt{1-\vec v^2}$ and $\vec u=\g\vec v$, and $\o_{\m\n}$ is a rank-2 skew tensor,
\begin{eqnarray}
\label{tenvor}
\o_{\m\n}=\pt_\m u_\n-\pt_\n u_\m,
\end{eqnarray}
which we will call the kinematic vorticity tensor. Hereafter, we will denote $\o^\m$ and $\o^{\m\n}$ as $\o_2^\m$ and $\o_2^{\m\n}$, respectively. It is worth writing down the components of $\o_2^\m$.
The spatial components are
\begin{eqnarray}
\label{omega-spa}
\vec\o_2=\g^2\vec\o_1+\g^2\vec v\times\pt_t\vec v,
\end{eqnarray}
and the time component is
\begin{eqnarray}
\o_2^0=\g^2\vec v\cdot\vec\o_1=\vec v\cdot\vec\o_2.
\end{eqnarray}
Thus, in the non-relativistic limit, $\o_2^\m\ra(h_{\rm f},\vec\o_1)$, as we expect.

In accord with the definitions of $\o_2^\m$ and $\o_{2\m\n}$, it seems natural to identify $\o_2^0$ as the relativistic helicity density and define the circulation along a closed contour line in spacetime as
\begin{eqnarray}
\oint u_\m dx^\m,
\end{eqnarray}
which, after using the Green's theorem, is transformed into the hypersurface integral of $\o_{2\m\n}$. However, such-defined circulation and the total helicity (the integral of $\o_2^0$ over space) are in general not conserved even for ideal barotropic fluid.

In order to maintain the circulation conservation and helicity conservation, other definitions of vorticity have been introduced~\cite{Lichnerowicz:1967,Rezzolla:2013}.
For example, if the fluid does not carry any conserved charge, one can define the vorticity tensor as~\cite{Florkowski:2013,Becattini:2015ska},
\begin{eqnarray}
\O_{\m\n}=\pt_\m(Tu_\n)-\pt_\n(Tu_\m),
\end{eqnarray}
where $T$ is temperature.
The corresponding circulation along a closed contour line in spacetime is defined by
\begin{eqnarray}
\oint T u_\m dx^\m.
\end{eqnarray}
By using the thermodynamic relations $d\ve=Tds$ and $dP=sdT$, the relativistic Euler equation for ideal fluid,
\begin{eqnarray}
\label{releuler}
(\ve+P)\frac{d}{d\t} u^\m=\nabla^\m P,
\end{eqnarray}
where $\ve$ and $P$ are the energy density and pressure, $d/d\t=u^\m\pt_\m$ is the proper time or comoving time derivative, and $\nabla_\m=\pt_\m-u_\m (d/d\t)$, can be deduced to
\begin{eqnarray}
\label{euler2}
\frac{d}{d\t}(Tu^\m)=\pt^\m T.
\end{eqnarray}
Thus one finds that for ideal fluid,
\begin{eqnarray}
\frac{d}{d\t}\oint Tu_\m dx^\m=\oint \pt_\m T dx^\m=0.
\end{eqnarray}
This is the relativistic Helmholtz-Kelvin theorem.

The vorticity (pseudo)vector corresponding to $\O_{\m\n}$ can be defined as
\begin{eqnarray}
\O^\m=\frac{1}{2}\e^{\m\n\r\s}Tu_\n \O_{\r\s}=T^2\o_2^\m.
\end{eqnarray}
It divergence reads
\begin{eqnarray}
\pt_\m\O^\m=\frac{1}{2}\e^{\m\n\r\s}\O_{\m\n} \O_{\r\s}=2\O^\m\O_{\m\n}u^\n,
\end{eqnarray}
which vanishes for ideal fluid upon noticing that \eq{euler2} can be rewritten as
\begin{eqnarray}
\O_{\m\n}u^\n=0.
\end{eqnarray}
Therefore, the integral of $\O^0$ over space is conserved and we can identify $\O^0$ as the conserved helicity density. (In general, when the quantum effect is taken into account, the flow helicity could be converted to other helicities, like the magnetic helicity or helicity of the constitutive fermions, and thus is not conserved~\cite{Avdoshkin:2014gpa,Yamamoto:2015gzz}.)

If the fluid carries a conserved charge (e.g., the baryon number), one can define the vorticity tensor as~\cite{Lichnerowicz:1967,Rezzolla:2013}
\begin{eqnarray}
\label{enthapyvor}
{\tilde\O}_{\m\n}=\pt_\m( wu_\n)-\pt_\n(wu_\m),
\end{eqnarray}
where $w=(\ve+P)/n$ is the enthalpy per particle with $n$ being the density of the conserved charge. The circulation conservation and helicity conservation formulated by using ${\tilde\O}_{\m\n}$ is presented in Appendix \ref{vort-enthapy}. However, in the following numerical simulations, we will not consider ${\tilde\O}_{\m\n}$ because the quark-gluon plasma produced in relativistic heavy-ion collisions carries almost zero conserved charges.

\section {Setup of the numerical simulations}\label{setup}
\begin{figure}[!htb]
\begin{center}
\includegraphics[width=6cm]{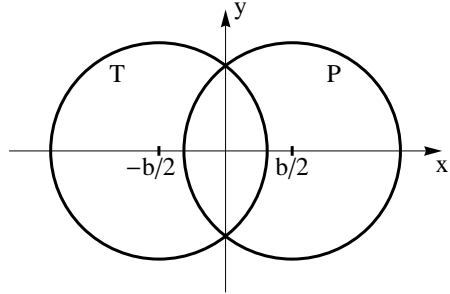}
\caption{Illustration of the heavy-ion collisions with impact parameter $b$.
Here ``T" stands for target and ``P" for projectile.}
\label{collision}
\end{center}
\end{figure}

In this section, we describe the general setup of our numerical simulations. The coordinate system of the heavy-ion collisions is illustrated in \fig{collision}. We choose the $z$ axis to be along the beam direction of the projectile, $x$ axis to be along the impact parameter ${\vec b}$ which points
from the target to the projectile, and $y$ axis to be perpendicular to
the reaction plane. The origin of the time axis, $t=0$, is set to the time when the two colliding nuclei overlap maximally.

We will focus mainly on the mid-rapidity region, but will also discuss how the vorticity varies with rapidity in \sect{subspat}. We will numerically compute the initial vorticity at proper time $\t=\t_0$ with the value of $\t_0$ will be given later. We will only briefly discuss the time evolution in \sect{sectime} based on analytical treatment of hydrodynamics. A full viscous hydrodynamic simulations is however beyond the scope of the present paper.

\subsection {Definition of the velocity field}\label{comp}
To compute the vorticity, we must first define the velocity field numerically. This can be achieved by introducing a smearing function $\F(x,x_i)$ where $x$ is the field point and $x_i$ is the coordinate of the $i$th particle. The effect of $\F(x,x_i)$ is to smear a physical quantity, e.g., energy or momentum, carried by the $i$th particle which locates at $x_i$ to another coordinate point $x$. Therefore, $\F(x,x_i)$ somehow represents the quantum wave packet of the $i$th particle. Having $\F(x,x_i)$, we can have two natural ways to define the velocity field for a given colliding event~\footnote{Note that one may introduce other ways to define the velocity field, for example, $v_{2'}^a(x)=\sum_ip^a_i\F(x,x_i)/\sum_i p^0_i \F(x,x_i)$ which, however, has less transparent physical meaning than $v^a_1$ and $v^a_2$. In fact, $v^a_{2'}$ is related to the energy momentum tensor by $v^a_{2'}=T^{0a}/T^{00}$ which in nonrelativistic limit is reduced to $v^a_2$; but in relativistic case, it represents neither the
velocity of energy flow nor the velocity of particle flow.},
\begin{eqnarray}
\label{defv1}
v^a_1(x)&=&\frac{1}{\sum_i \F(x,x_i)}\sum_i\frac{p^a_i}{p^0_i}\F(x,x_i),\\
\label{defv2}
v^a_2(x)&=&\frac{\sum_i p^a_i \F(x,x_i)}{\sum_i [p^0_i+(p_i^a)^2/p_i^0] \F(x,x_i)},
\end{eqnarray}
where $a=1, 2, 3$ is the spatial indices, $\bp_i$ and $p^0_i$ are the momentum and energy of the $i$th particle, and the summation is over all the particles. In our simulations, $p_i$ and $x_i$ in each event are generated by HIJING.

Now we clarify the physical meanings of $\bv_1$ and $\bv_2$. Let $f(x,p)$ be the particle distribution function. It is related to $\F(x,x_i)$ by
\begin{eqnarray}
f(x,p)=\frac{1}{{\cal N}}\sum_{i}(2\p)^3\d^{(3)}[\bp-\bp_i(t)] \F(x,x_i),
\end{eqnarray}
with ${\cal N}=\int d^3\bx \F(x,x_i)$ a normalization factor. Then the energy-momentum tensor and particle number current are given by
\begin{eqnarray}
T^{\m\n}(x)&=&\int \frac{d^3\bp}{(2\p)^3} \frac{p^\m p^\n}{p^0} f(x,p)=\frac{1}{\cal N}\sum_i\frac{p^\m_i p^\n_i}{p_i^0}\F(x,x_i),\\
J^{\m}(x)&=&\int \frac{d^3\bp}{(2\p)^3} \frac{p^\m}{p^0} f(x,p)=\frac{1}{\cal N}\sum_i\frac{p^\m_i}{p_i^0}\F(x,x_i).
\end{eqnarray}
These give
\begin{eqnarray}
T^{0a}&=&\frac{1}{\cal N}\sum_i p^a_i\F(x,x_i),\non
T^{00}&=&\frac{1}{\cal N}\sum_i p^0_i\F(x,x_i),\non
T^{ab}&=&\frac{1}{\cal N}\sum_i \frac{p^a_ip^b_i}{p_i^0}\F(x,x_i),\non
J^0&=&\frac{1}{\cal N}\sum_i\F(x,x_i),\non
J^a&=&\frac{1}{\cal N}\sum_i\frac{p^a_i}{p_i^0}\F(x,x_i).
\end{eqnarray}
Thus we can identify that
\begin{eqnarray}
v^a_1&=&\frac{J^a}{J^0},\\
v^a_2&=&\frac{T^{0a}}{T^{00}+T^{aa}},
\end{eqnarray}
that is, $\bv_1$ is the velocity of the particle flow associated with $J^\m$ and $\bv_2$ is the velocity of the energy flow (see Appendix \ref{append-coo}). We note that the frame in which the flow velocity is chosen to be $\bv_1$ is usually called Eckart frame (more precisely, Eckart frame requires $J^\m$ to be associated with a conserved charge which is not the case for a gluon-dominated partonic matter) while the frame in which the flow velocity is $\bv_2$ is usually called Landau frame~\cite{DeGroot:1980dk}.

Different choice for the smearing function $\F(x,x_i)$ gives different result for the velocity. In our computations, we choose a smearing function whose functional form at $\t=\t_0$ is a Gaussian~\cite{Pang:2012he},
\begin{eqnarray}
\label{smear2}
\F_{\rm G}(x,x_i)=\frac{K}{\t_0\sqrt{2\p\s_\w^2}2\p\s_r^2}\exp{\ls-\frac{(x-x_i)^2+(y-y_i)^2}{2\s_r^2}-\frac{(\w-\w_i)^2}{2\s_\w^2}\rs},
\end{eqnarray}
where $\s_r$ and $\s_\w$ are two width parameters and $K$ is a scale factor. The spacetime rapidity and proper time are define by $\w=(1/2)\ln[(t+z)/(t-z)]$ and $\t=\sqrt{t^2-z^2}$. This kind of smearing function has been widely used in hydrodynamic simulations, e.g., in Refs.~\cite{Pang:2012he,Hirano:2012kj}. The parameters that we will use are the following. For RHIC Au + Au collisions at $\sqrt{s}=200$ GeV: $\s_r=0.6$ fm, $\s_\w=0.6$, $K=1.45$, and $\t_0=0.4$ fm; For LHC Pb + Pb collisions at $\sqrt{s}=2.76$ TeV: $\s_r=0.6$ fm, $\s_\w=0.6$, $K=1.6$, and $\t_0=0.2$ fm. The initial energy momentum tensor obtained by using these parameters can fit the experimental data quite well after the viscous hydrodynamic evolution~\cite{Pang:2012he}. We note that the parameters $K$ and $\t_0$ in $\F_{\rm G}$ do not change the velocity because they cancel out in \eq{defv1} and \eq{defv2}; however, they do influence the values of energy density and temperature.

We in Appendix \ref{newsmearing} discuss another smearing function $\F_\D(x,x_i)$ and give the numerical result for velocity field computed by using $\F_\D(x,x_i)$.

After performing the event average, we have
\begin{eqnarray}
\label{v-12}
\lan v_1^a\ran(x)&\equiv&\frac{1}{N_e}\sum_ev_1^a(x),\\
\label{v-122}
\lan v^a_2\ran(x)&\equiv&\frac{1}{N_e}\sum_e v_2^a(x),
\end{eqnarray}
where $\lan\cdots\ran$ denotes average over events, $N_e$ is the total number of events, and the summation of $e$ is over all the events. For the purpose of numerical check, we will also compute the following quantities:
\begin{eqnarray}
\label{v-12p}
\lan v_3^a\ran(x)&\equiv&\frac{\sum_e\sum_{i\in e}(p_i^a/p^0)\F(x,x_i)}{\sum_e\sum_{i\in e}\F(x,x_i)},\\
\label{v-12p2}
\lan v_4^a\ran(x)&\equiv&\frac{\sum_e\sum_{i\in e}p^a_i\F(x,x_i)}{\sum_e\sum_{i\in e} [p^0+(p^a_i)^2/p^0]\F(x,x_i)}.
\end{eqnarray}
We note that $\lan\bv_1\ran$ and $\lan\bv_2\ran$ are event-averaged $\bv_1$ and $\bv_2$, while
$\lan\bv_3\ran$ and $\lan\bv_4\ran$ can be considered as first accumulating $N_e$ events into one event and then
calculating $\bv_1$ and $\bv_2$ of that event; $\lan\bv_3\ran$ and $\lan\bv_4\ran$ cannot be defined on event-by-event basis.

\subsection {Definition of the vorticity and helicity fields}\label{secvordef}
Once the specific definition of the velocity field is given in numerical setup, the vorticity is calculated according to
\begin{eqnarray}
\label{define-vor1}
{\bm\o}_{1}&=&{\bm\nabla}\times\bv,\\
\label{define-vor2}
{\bm\o}_{2}&=&\g^2{\bm\nabla}\times\bv.
\end{eqnarray}
Note that we have neglected the $O(v^2)$ term $\g^2\bv\times\pt_t\bv$ in ${\bm\o}_2$ (see \eq{omega-spa}) because, as we will show, the velocity is small in the central overlapping region that we are most interested in. We will also compute various helicity densities:
\begin{eqnarray}
h_{\rm f}&=&\bv\cdot\vec \o_{1},\\
\o_2^0&=&\bv\cdot\vec \o_2,\\
\O^0&=&T^2\bv\cdot\vec \o_2.
\end{eqnarray}

\section {Global angular momentum and local shear flow}\label{globa}
\begin{figure}[!htb]
\begin{center}
\includegraphics[width=7cm]{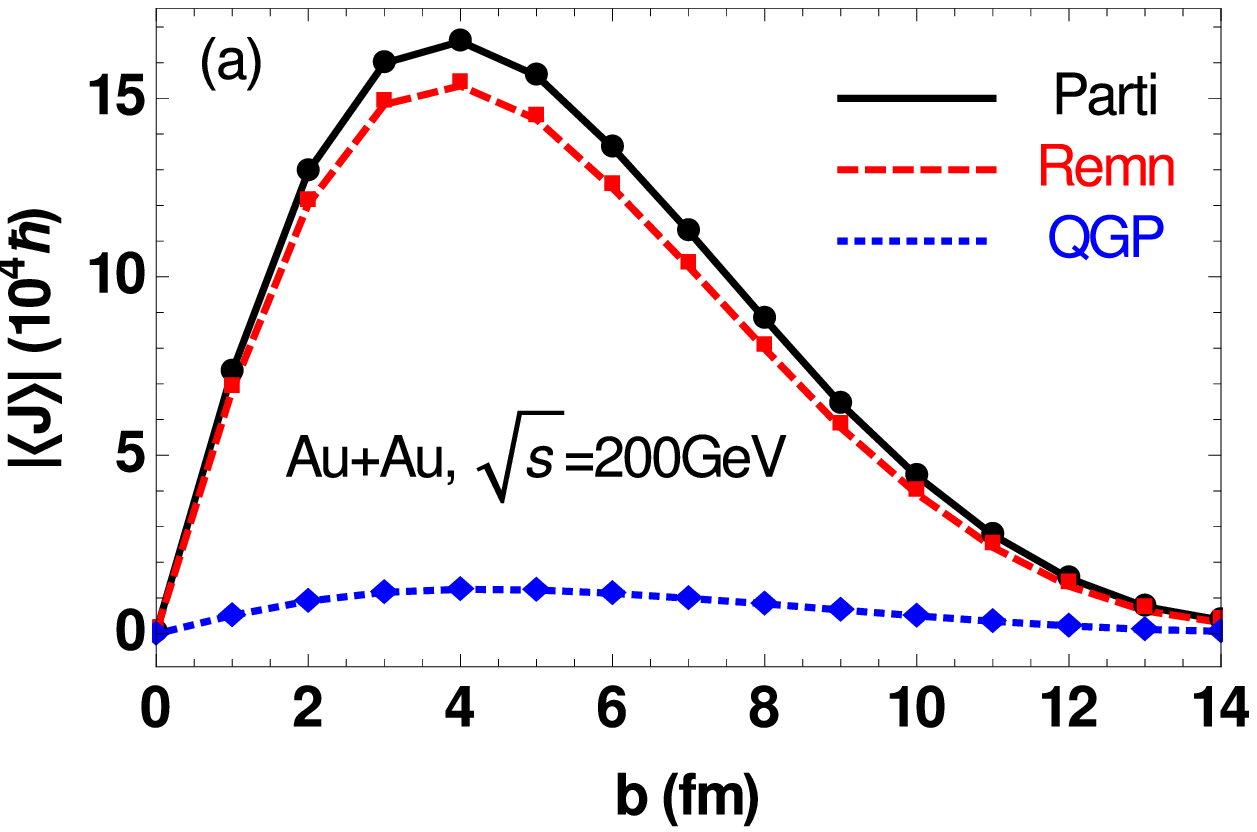}
\includegraphics[width=7cm]{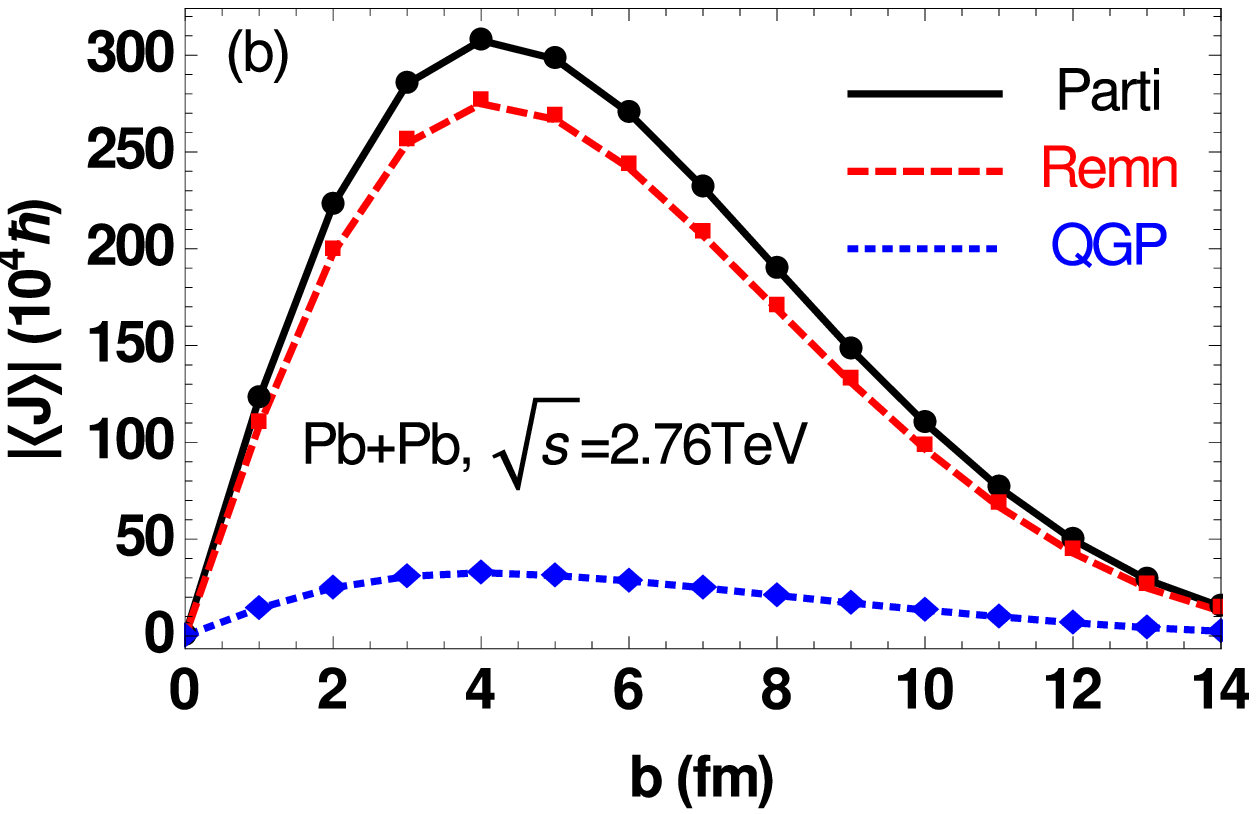}
\caption{The impact parameter dependence of the angular momenta of the participents (black), the remnants (red), and the QGP (blue) with respect to the collision center $\bx=\vec0$ for RHIC Au + Au collsions (panel (a)) and LHC Pb + Pb collisions (panel (b)). The sum of the angular momenta of the remnants and the QGP is equal to the angular momentum of the participants.}
\label{fig-angmom}
\end{center}
\end{figure}
As we have discussed in the Sec.~\ref{intro}, in a non-central heavy-ion collision with impact parameter $b$, the total angular momentum $J_0$ of the two nuclei with respect to the collision center is roughly given by $Ab\sqrt{s}/2$. After the collision, a fraction of $J_0$ is carried away by the spectators which fly rapidly apart from the collision region, the remained fraction of $J_0$ is carried by the remnant nucleons (the wounded participants with large longitudinal momenta) as well as the produced QGP. In \fig{fig-angmom} we show our numerical simulations for the event-averaged angular momenta carried by the participants, the remnants, and the QGP for Au + Au collisions at $\sqrt{s}=200$ GeV and for Pb + Pb collisions at $\sqrt{s}=2.76$ TeV. The results are obtained by averaging over $10^5$ events. In calculating the angular momentum, we use the formula
\begin{eqnarray}
J=\int d^3\bx(zT^{0x}-xT^{0z}).
\end{eqnarray}
Our results for the Au + Au collisions are qualitatively consistent with previous studies in Refs.~\cite{Gao:2007bc,Becattini:2007sr,Vovchenko:2013viu,Becattini:2015ska}. We find that about $10\%$ of the angular momentum of the total participants is retained by the QGP at mid-centrality region.

\begin{figure}[!htb]
\begin{center}
\includegraphics[width=7cm]{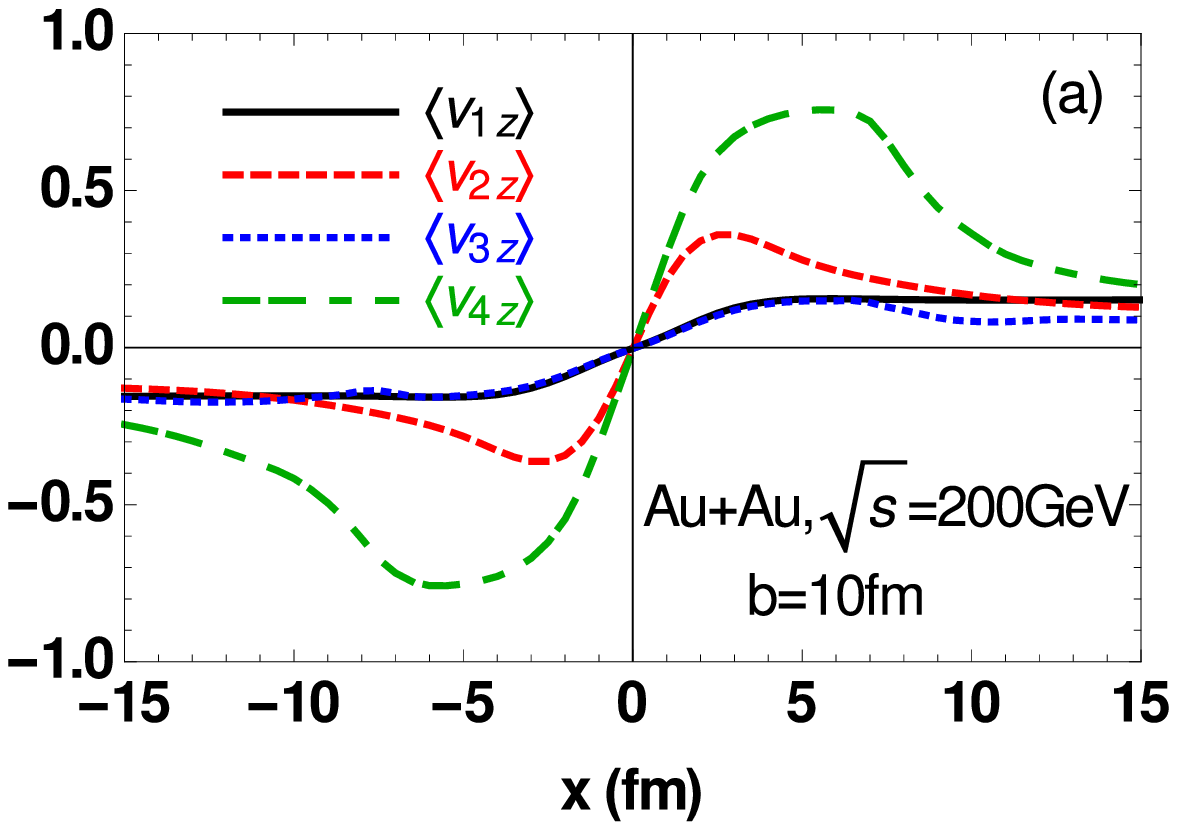}
\includegraphics[width=7cm]{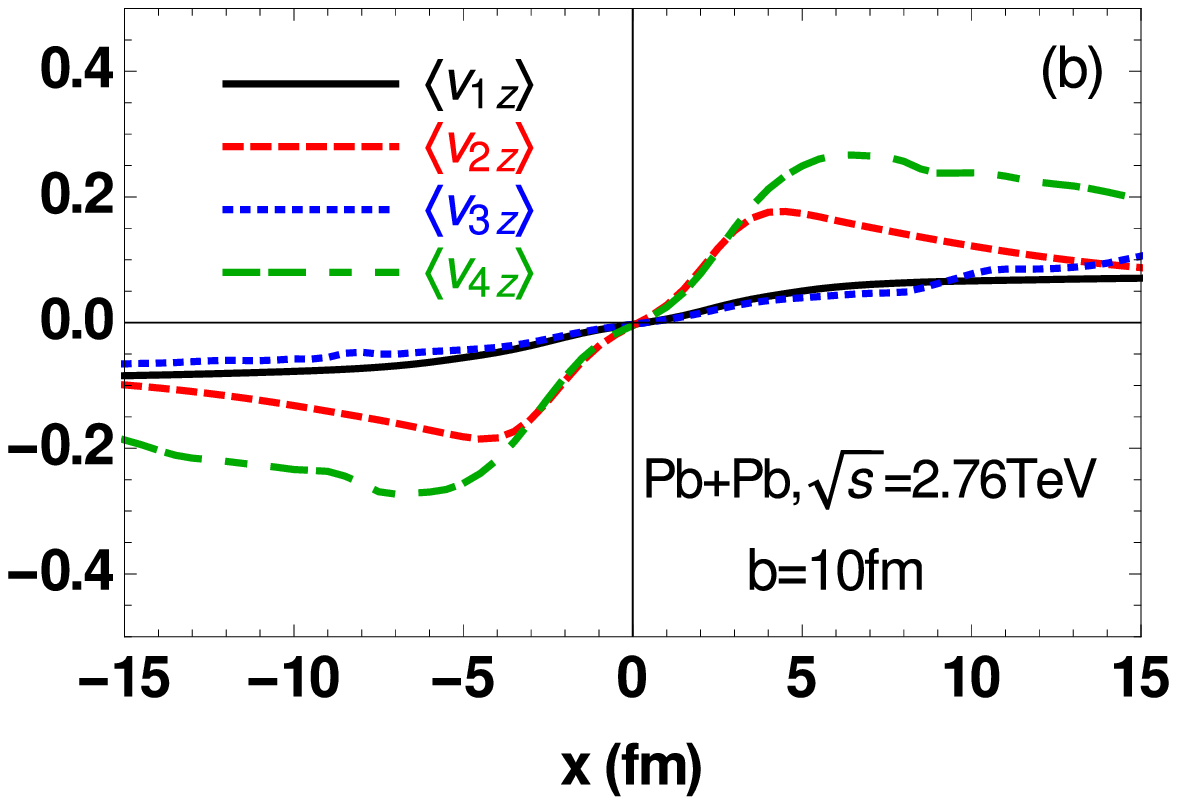}
\caption{The event-averaged longitudinal velocity profile at zero rapidity for RHIC (panel (a)) and LHC (panel (b)). Different curves correspond to different definitions of the event-averaged velocity, see \eq{v-12} $-$ \eq{v-12p2}.}
\label{fig-vz-prof}
\end{center}
\end{figure}

Such a global angular momentum of the QGP manifests itself mainly in the form of local fluid shear rather than a global rigid rotation. Our numerical result for the event-averaged longitudinal shear flow profile at zero rapidity and $b=10$ fm is presented in \fig{fig-vz-prof} (a) for Au + Au collisions at $\sqrt{s}=200$ GeV and in \fig{fig-vz-prof} (b) for Pb + Pb collisions at $\sqrt{s}=2.76$ TeV, where we show $\lan v_z\ran (x)$ as a function of the transverse coordinate $x$ for four different definitions of $\lan v_z\ran$ according to \eq{v-12} $-$ \eq{v-12p2}. We emphasize that due to the use of the Gaussian smearing function $\F_G$, $\lan v_z\ran (x)$ remains finite even for very large values of $x$ where the QGP is not expected to realistically exist. The simulation is more sensible for smaller values of $x$ so that we will mainly concentrate on the region near the collision center in the following discussions.

It is obvious that for given impact parameter, the angular momentum in Pb + Pb collisions at $\sqrt{s}=2.76$ TeV is larger than that in Au + Au collisions at $\sqrt{s}=200$ GeV; however, \fig{fig-vz-prof} gives that the longitudinal velocity at $\w=0$ for Pb + Pb collisions is smaller than that for Au + Au collisions at the same impact parameter. As we checked that this is partially attributable to the fact that for larger collision energy the moment of inertia of the partonic system is also larger and partially attributable to that for larger collision energy the large-rapidity particles contribute a larger fraction of the total angular momentum. We will present further discussion on this issue when we discuss the collision energy dependence of vorticity in Sec.~\ref{subene}. The main information from \fig{fig-vz-prof} is that near the center ($x=0$) of the overlapping region, $\lan v_z\ran(x)$ grows with $x$ and thus has a finite shear that naturally suggests a finite local vorticity perpendicular to the reaction plane which we
now study in detail.

\section {Numerical results for vorticity}\label{resvor}
\subsection {Impact parameter dependence}\label{subimp}
We begin by studying how the vorticity at $\t=\t_0$ depends on the impact parameter $b$. For each event, we compute the vorticity at $\w=0$ averaged over the overlapping region in the transverse plane according to
\begin{eqnarray}
\label{spa-vor1}
\bar{\vec\o}\equiv\frac{\int d^2\bx_\perp n(\bx_\perp)\vec\o(\bx_\perp)}{\int d^2\bx_\perp n(\bx_\perp)}
\end{eqnarray}
if the vorticity is computed by using the particle flow velocity $\bv_1$, or
\begin{eqnarray}
\label{spa-vor2}
\bar{\vec\o}\equiv\frac{\int d^2\bx_\perp \ve(\bx_\perp)\vec\o(\bx_\perp)}{\int d^2\bx_\perp \ve(\bx_\perp)}
\end{eqnarray}
if the vorticity is computed by using the energy flow velocity $\bv_2$. In \eq{spa-vor1} and \eq{spa-vor2}, $\vec\o$ is representative of either $\vec\o_1$ or $\vec\o_2$, $n(\bx_\perp)$ and $\ve(\bx_\perp)$ are the parton number density and energy density obtained in HIJING, and $\bx_\perp$ is the coordinate in the transverse plane. Such space-averaged vorticity more appropriately reflects the strength of the vorticity acting on the whole overlapping region. We then average $\bar{\vec\o}$ over $10^5$ events to obtain the event-averaged space-averaged vorticity, $\lan\bar{\vec\o}\ran$, which we will call the double-averaged vorticity.

\begin{figure}[!htb]
\begin{center}
\includegraphics[width=7cm]{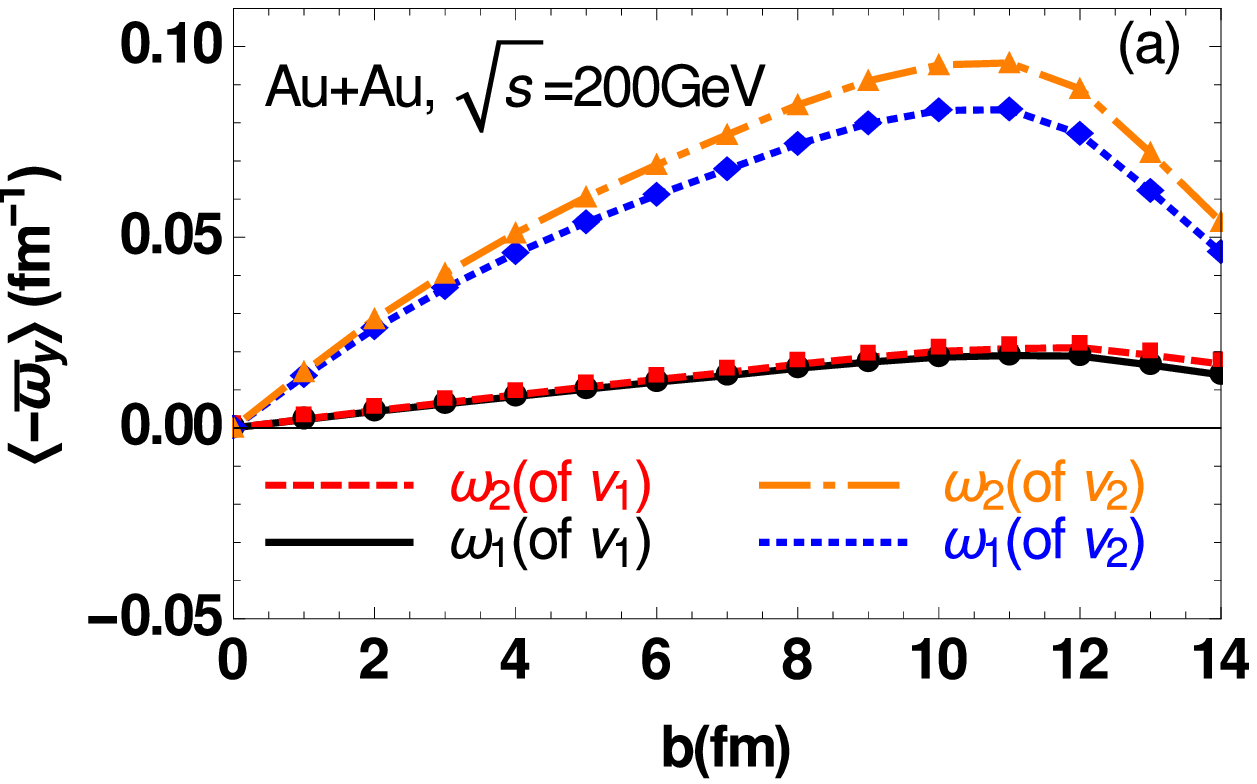}
\includegraphics[width=7cm]{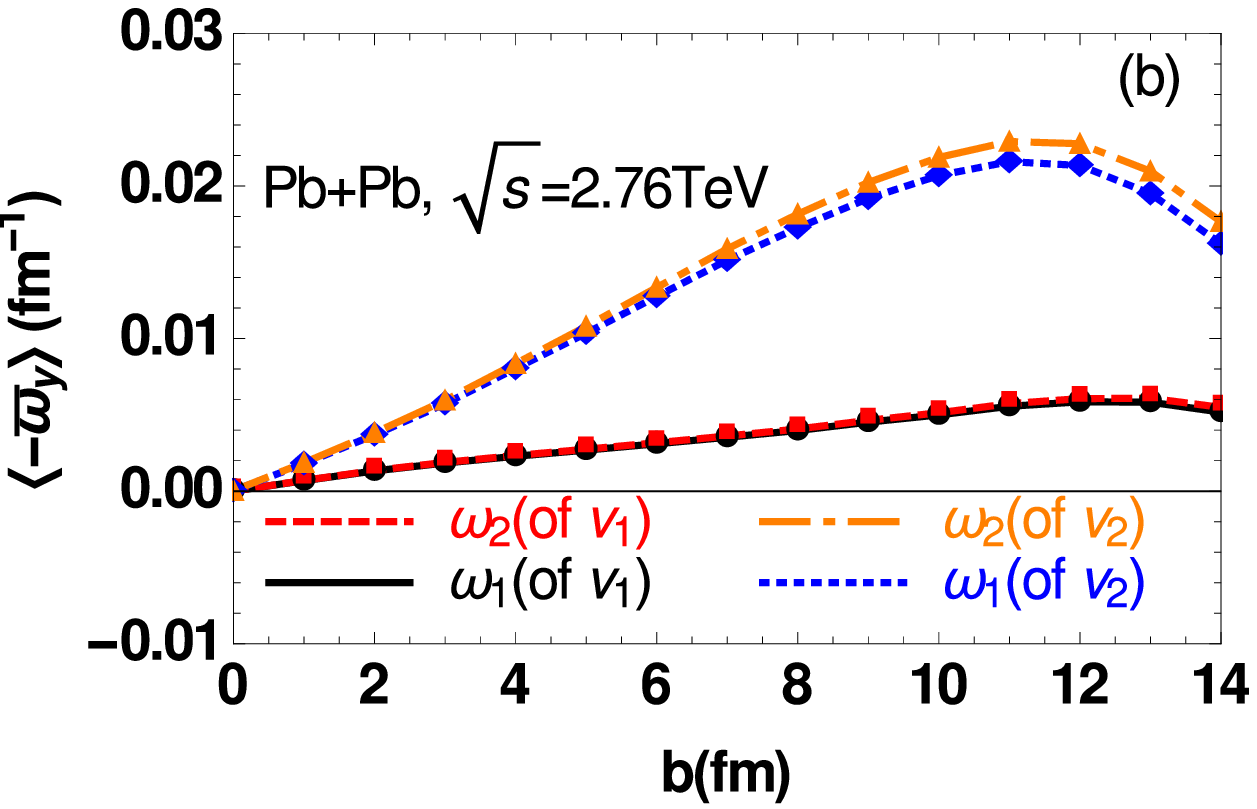}
\caption{The double-averaged vorticity at $\t=\t_0$ and $\w=0$ for RHIC Au + Au collisions at $\sqrt{s}=200$ GeV (panel (a)) and LHC Pb + Pb collisions at $\sqrt{s}=2.76$ TeV (panel (b)). Only the $y$-component of the vorticity is sizable, other components are negligibly small and are not drawn. Different curves correspond to different definitions of vorticity and velocity.}
\label{fig-vor-b-200}
\end{center}
\end{figure}

In \fig{fig-vor-b-200}, we show the $y$-component of the double-averaged vorticities $\lan\bar{\vec\o}_{1}(\t_0,\w=0)\ran$ and $\lan\bar{\vec\o}_2(\t_0,\w=0)\ran$ for RHIC Au + Au collisions at $\sqrt{s}=200$ GeV and LHC Pb + Pb collisions at $\sqrt{s}=2.76$ TeV. We have checked that after the event average the $x$ and $z$ components of the vorticity are vanishing, as we expected from the left-right symmetry of the colliding geometry.

First, we notice that the magnitude of the vorticity generated in either RHIC or LHC is big. For example, $\lan\bar{\o}_{y}\ran$ of the energy flow $\bv_2$ at $b=10$ fm is about $10^{21} s^{-1}$ or $20$ MeV for RHIC Au + Au collisions at $\sqrt{s}=200$ GeV. According to \eq{cve2}, the strength of axial CVE is roughly proportional to $T^2\lan\bar{\o}_{y}\ran\sim10^6$ MeV$^3$ if we plausibly assume $T\sim 300$ MeV. This is competitive to the strength of CSE in RHIC which is proportional to $\m eB_y$ if the vector chemical potential is about $\m\sim 10$ MeV and the magnetic field $eB_y\sim 5 m_\p^2$.  Second, the vorticity of energy flow is generally larger than the vorticity of particle flow, in consistence with the longitudinal velocity profile, \fig{fig-vz-prof}. Third, as $b$ grows from zero, $\lan\bar{\o}_y\ran$ first increases and reaches its maximum value at about $b\simeq 2R_A$, after that the two nuclei are essentially not colliding and $\lan\bar{\o}_y\ran$ drops. This behavior is consistent with the fact that the angular momentum of the QGP shows a similar non-monotonous feature.

\begin{figure}[!htb]
\begin{center}
\includegraphics[width=7cm]{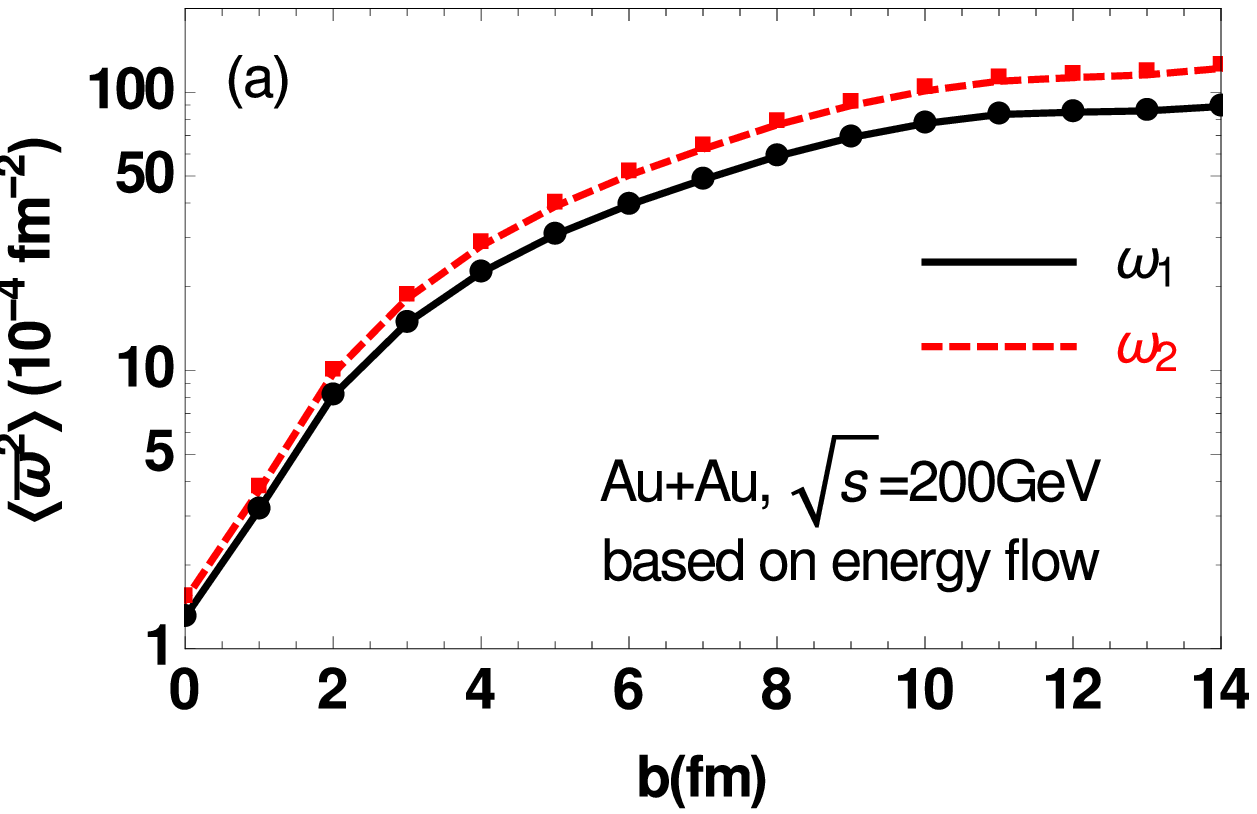}
\includegraphics[width=7cm]{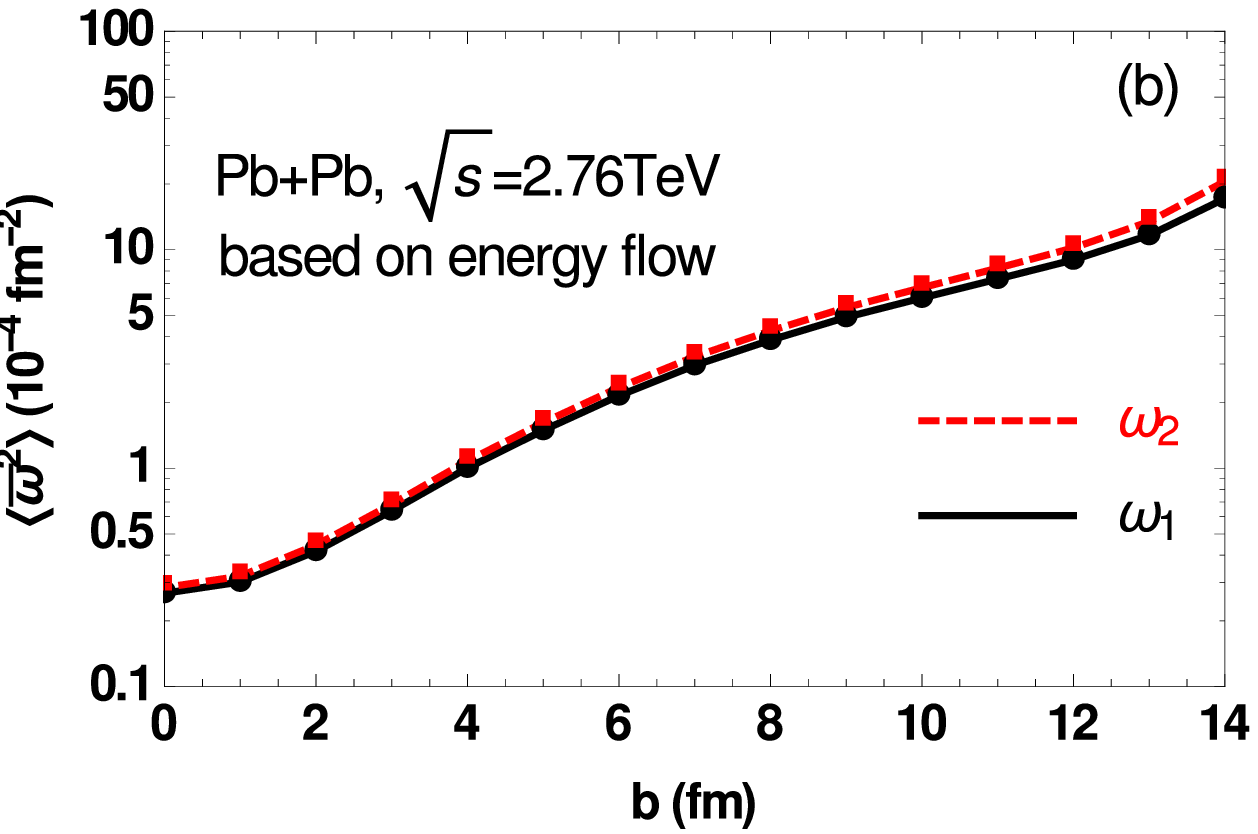}
\caption{The vorticity squared of the energy flow at $\t=\t_0$ and $\w=0$ for RHIC Au + Au collisions at $\sqrt{s}=200$ GeV (panel (a)) and LHC Pb + Pb collisions at $\sqrt{s}=2.76$ TeV (panel (b)). Different curves correspond to different definitions of the vorticity.}
\label{fig-vor-squ}
\end{center}
\end{figure}

Although the $x$ and $z$ components of the vorticity vanish after averaging over many events, their magnitudes for each event can be finite due to the fluctuation of the nucleon positions in the nucleus. Such event-by-event fluctuation of vorticity magnitude is characterized by $\lan\bar{\vec\o}^2\ran$ which we show in \fig{fig-vor-squ} where the vorticity is calculated based on the energy flow $\bv_2$. If there is no event-by-event fluctuation, $\lan\bar{\vec\o}^2\ran$ should be equal to $\lan\bar{\o}_y\ran^2$. But by comparing \fig{fig-vor-squ} with \fig{fig-vor-b-200}, we observe clearly that $\lan\bar{\vec\o}^2\ran>\lan\bar{\o}_y\ran^2$. This is most evident for $b=0$ where $\lan\bar{\o}_y\ran=0$ but $\lan\bar{\vec\o}^2\ran$ is still finite. Another feature that \fig{fig-vor-squ} exhibits is that for $b\gtrsim 2R_A$, unlike $\lan\bar{\o}_y\ran$, $\lan\bar{\vec\o}^2\ran$ does not drop, which indicates that the event-by-event fluctuation of the vorticity is stronger for larger $b$.

\subsection {Collision energy dependence}\label{subene}
\begin{figure}[!htb]
\begin{center}
\includegraphics[width=7cm]{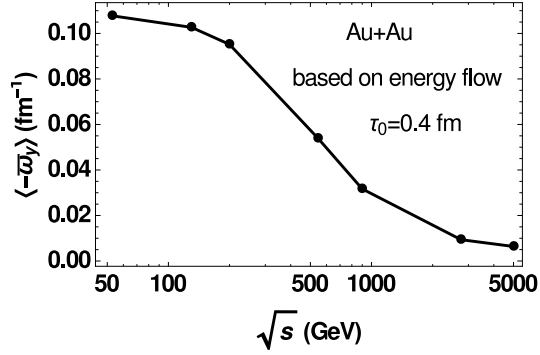}
\caption{The collision energy dependence of the double-averaged vorticity $\lan\bar{\o}_{2y}\ran$ at fixed $\t=0.4$ fm and $\w=0$ computed by using the energy flow velocity $\bv_2$. }
\label{fig-vor-ene}
\end{center}
\end{figure}
From \fig{fig-vor-b-200}, one can observe that, for given impact parameter, the vorticity for Pb + Pb collisions at $\sqrt{s}=2.76$ TeV is smaller than that for Au + Au collisions at $\sqrt{s}=200$ GeV. This suggests that the vorticity $\lan\bar{\o}_y\ran$ decreases when the collision energy increases. We thus perform the numerical simulation for $\lan\bar{\o}_{2y}\ran$ of $\bv_2$ at fixed $\t=0.4$ fm and $\w=0$ in Au + Au collisions with $b=10$ fm at different $\sqrt{s}$. The result is drawn in \fig{fig-vor-ene} which shows clearly a decreasing vorticity as $\sqrt{s}$ increases. We note that similar feature was also observed in Ref.~\cite{Jiang:2016woz} where the event-averaged moment-of-inertia weighted vorcity was computed. At first sight, this behavior appears to contradict the collision energy dependence of the angular momentum of QGP which increases with $\sqrt{s}$ (see \fig{fig-angmom}), because the vorticity measures the local angular velocity of the fluid, thus the whole angular momentum of QGP would be
roughly $\vec J\sim\int d^3\bx I(\bx)\vec \o(\bx)$ where $I(\bx)\sim [\bx^2-(\bx\cdot \hat{\vec\o})^2]\ve(\bx)$ is the moment of inertia density. However, as noticed in Ref.~\cite{Jiang:2016woz}, with increasing collision energy, the moment of inertia grows more rapidly than the increasing of the total angular momentum of QGP, and thus makes the vorticity decrease. Furthermore, as we will show in next subsection, with increasing collision energy, the vorticity becomes more and more peaked at finite rapidity and thus the vorticity at $\w=0$ is relatively weakened. This reflects the fact that at higher collision energy, the system at the mid-rapidity region behaves to be closer to the Bjorken boost invariant picture and thus allows smaller vorticity.

\subsection {Spatial distribution of vorticity}\label{subspat}
\begin{figure}[!htb]
\begin{center}
\includegraphics[width=7cm]{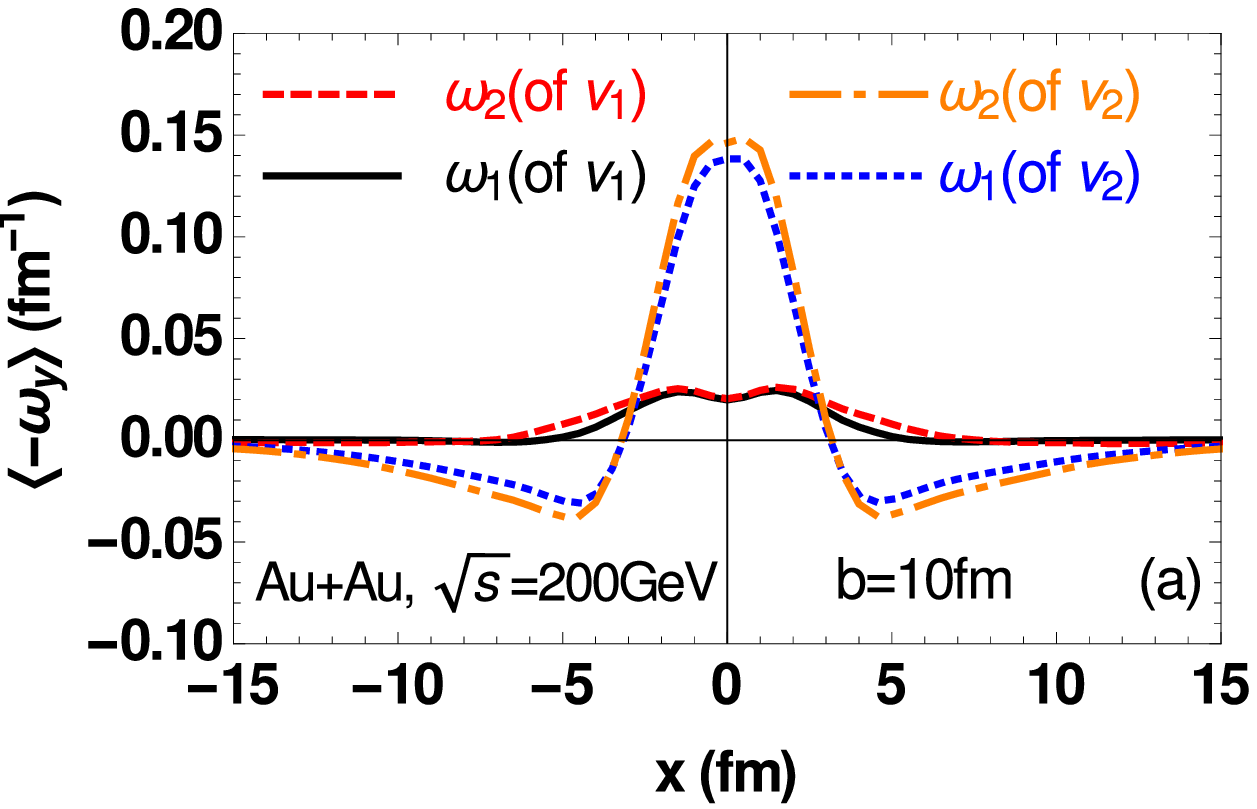}
\includegraphics[width=7cm]{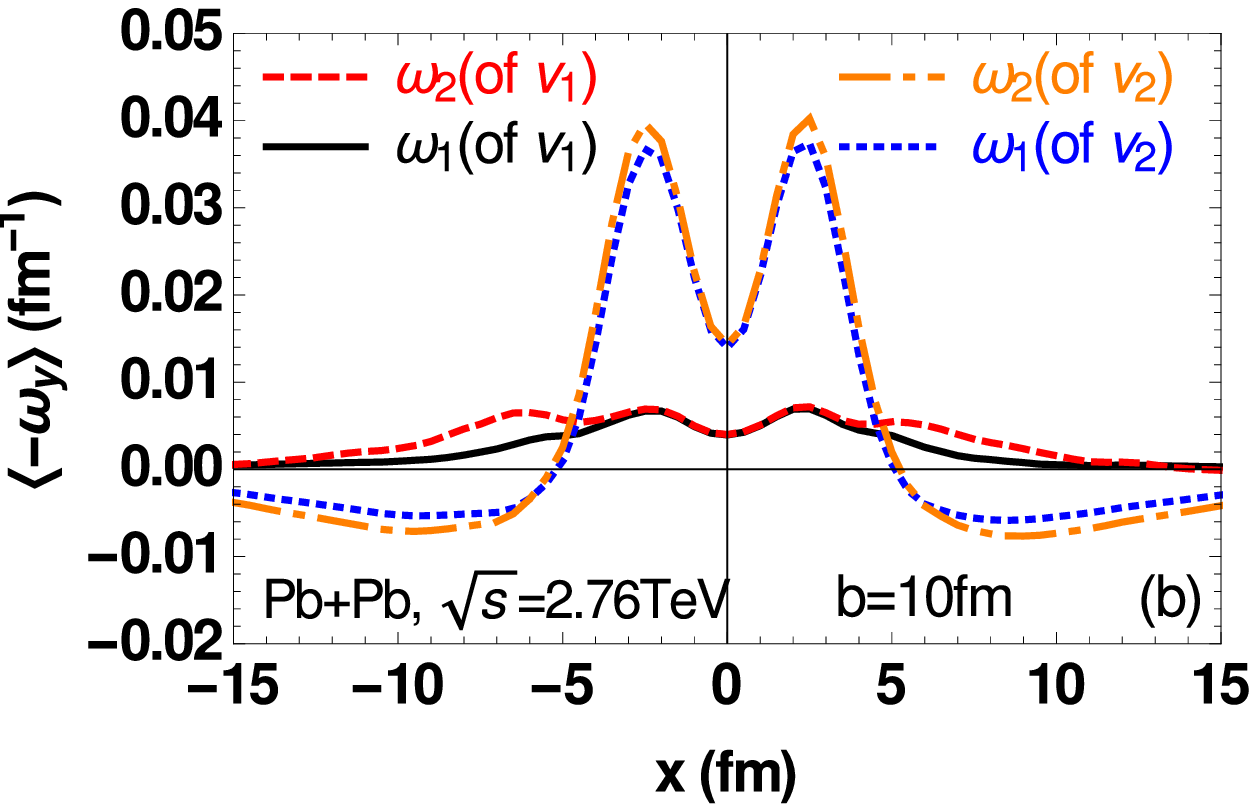}
\caption{The event-averaged vorticity as a function of $x$ for RHIC Au + Au collisions at $\sqrt{s}=200$ GeV (panel (a)) and LHC Pb + Pb collisions at $\sqrt{s}=2.76$ TeV (panel (b)). Different curves represent different definitions of vorticity (see \eq{define-vor1} and \eq{define-vor2}) based on different definitions of the velocity field (see \eq{defv1} and \eq{defv2}).}
\label{fig-vor-prof-200}
\end{center}
\end{figure}
\begin{figure}[!htb]
\begin{center}
\includegraphics[width=7cm]{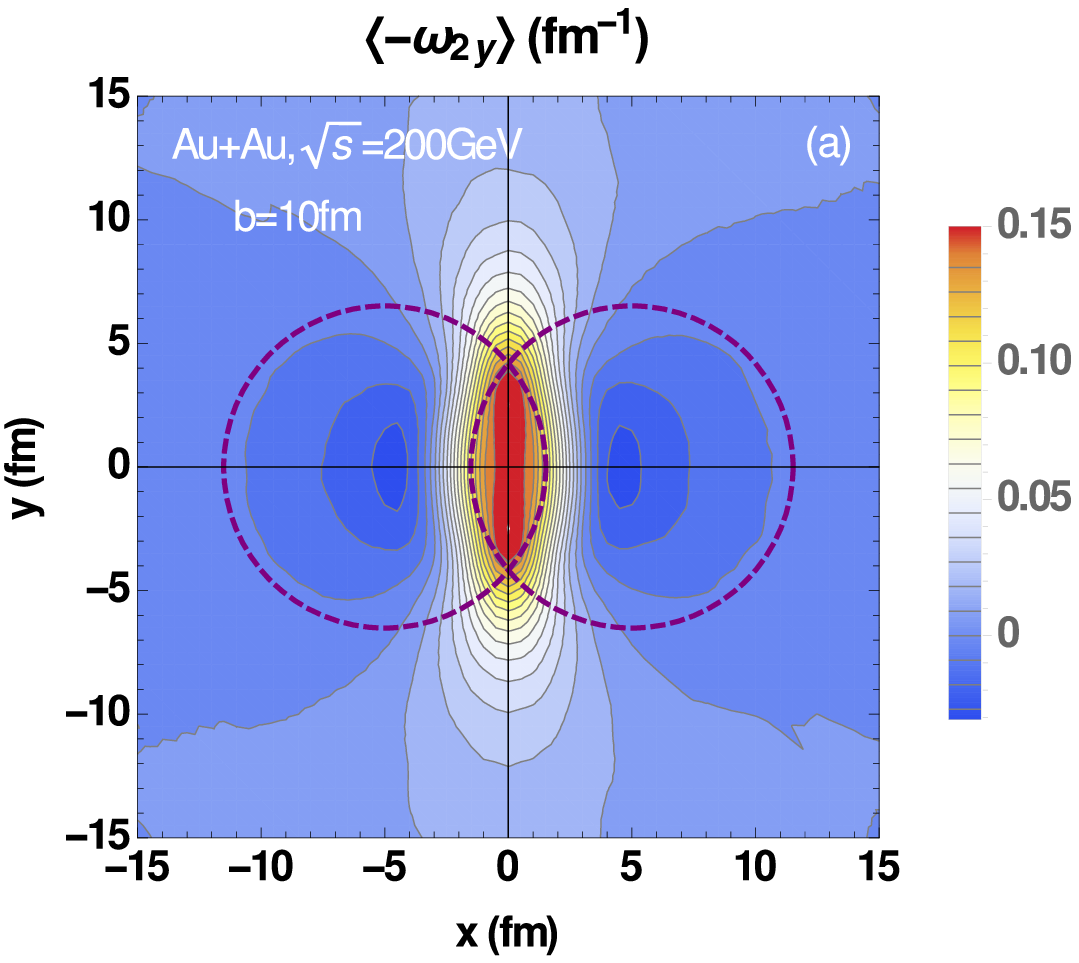}
\includegraphics[width=7cm]{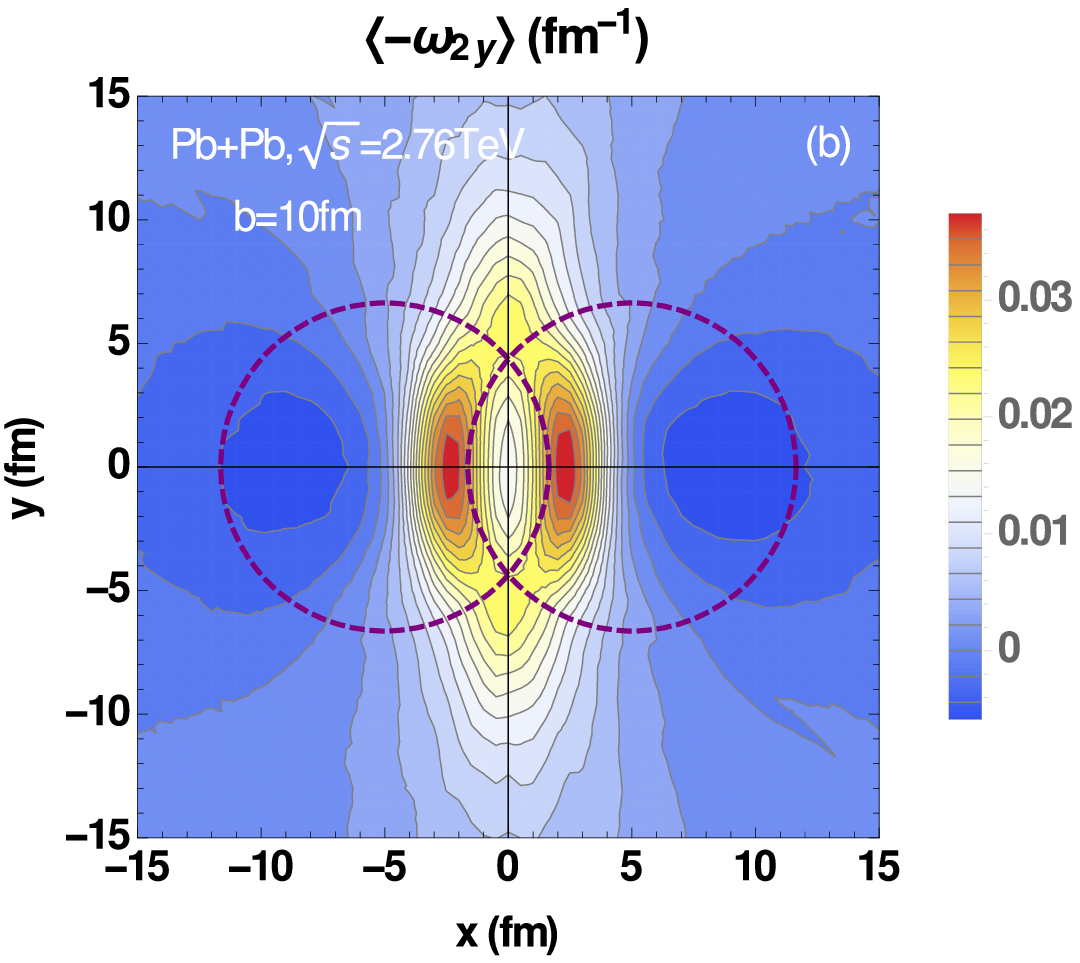}
\caption{The spatial distribution of the event-averaged vorticity, $\lan\o_{2y}\ran$, in the transverse plane for RHIC Au + Au collisions at $\sqrt{s}=200$ GeV (panel (a)) and LHC Pb + Pb collisions at $\sqrt{s}=2.76$ TeV (panel (b)).}
\label{fig-vor-spa-200}
\end{center}
\end{figure}

We show in \fig{fig-vor-prof-200} the event-averaged vorticity as a function of $x$ (the coordinate in the impact parameter direction). The full distribution of the vorticity (we present only $\lan\o_{2y}\ran$ of $\bv_2$ as an example) in the transverse plane (the $x-y$ plane) is shown in \fig{fig-vor-spa-200}. Obviously, the vorticity distributes in the transverse plane very inhomogeneously. From \fig{fig-vor-prof-200} and \fig{fig-vor-spa-200} we observe three remarkable features. (1) From \fig{fig-vor-spa-200} we notice that $\lan\o_{2y}\ran$ varies more steeply along the $x$ direction than along the $y$ direction in consistence with the elliptic shape of the overlapping region. (2) From \fig{fig-vor-prof-200}, one finds that the magnitude of $\lan\o_y\ran$ reaches its maximum value not at the center ($x=0$) but at a position $x_p$ which becomes larger for higher collision energy and finally becomes well localized around the outer boundary of the overlapping region (we have checked this for collision
energy other than $200$ GeV and $2.76$ TeV). Note that although in \fig{fig-vor-prof-200} and \fig{fig-vor-spa-200} it seems that the vorticity of $\bv_2$ at $\sqrt{s}=200$ GeV peaks at $x=0$, this is not the case as we have checked at higher resolution near $x=0$. (3) The vorticity drops steeply for $x$ larger than $x_p$. The vorticity of energy flow even shows a flipping of direction at $x\gg x_p$ which is due to the drop of $\lan v_{2z}\ran(x)$ at large $x$ as shown in \fig{fig-vz-prof}. The last two features are closely related to the boundary of the overlapping region and thus can be called a corona effect for vorticity which reflects the fact that near in the boundary layers the velocity field varies severely.

\begin{figure}[!htb]
\begin{center}
\includegraphics[width=7cm]{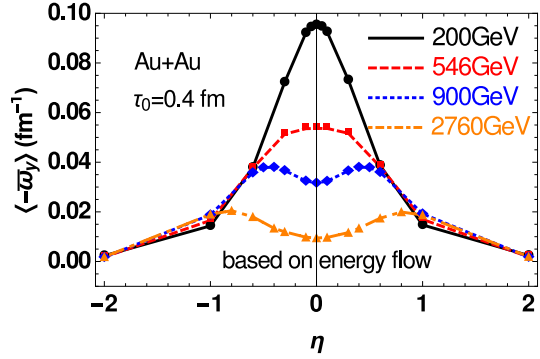}
\caption{The double-averaged vorticity $\lan\bar{\o}_{2y}\ran$ in Au + Au collisions calculated based on $\bv_2$ as a function of spacetime rapidity at various collision energies. The proper time is fixed to be $\t_0=0.4$ fm.}
\label{fig-vor-rap}
\end{center}
\end{figure}
In \fig{fig-vor-rap} we show the double-averaged vorticity $\lan\bar{\o}_{2y}\ran$ of energy flow as a function of the spacetime rapidity in Au + Au collisions for various collision energies. We find that for collision energy $\sqrt{s}\lesssim 550$ GeV $\lan\bar{\o}_{2y}\ran$ peaks at zero rapidity (for our best resolution) while for $\sqrt{s}>550$ GeV it peaks at a finite rapidity which increases as $\sqrt{s}$ grows. This feature may be understood by noticing that for fixed proper time, the boundary of the collision region in the $\w$ direction is increasing with $\sqrt{s}$; thus the appearance of the finite-rapidity peak may be also considered as a corona effect in $\w$ direction.

\subsection {Spatial distribution of helicity}\label{subhel}
\begin{figure}[!htb]
\begin{center}
\includegraphics[width=7cm]{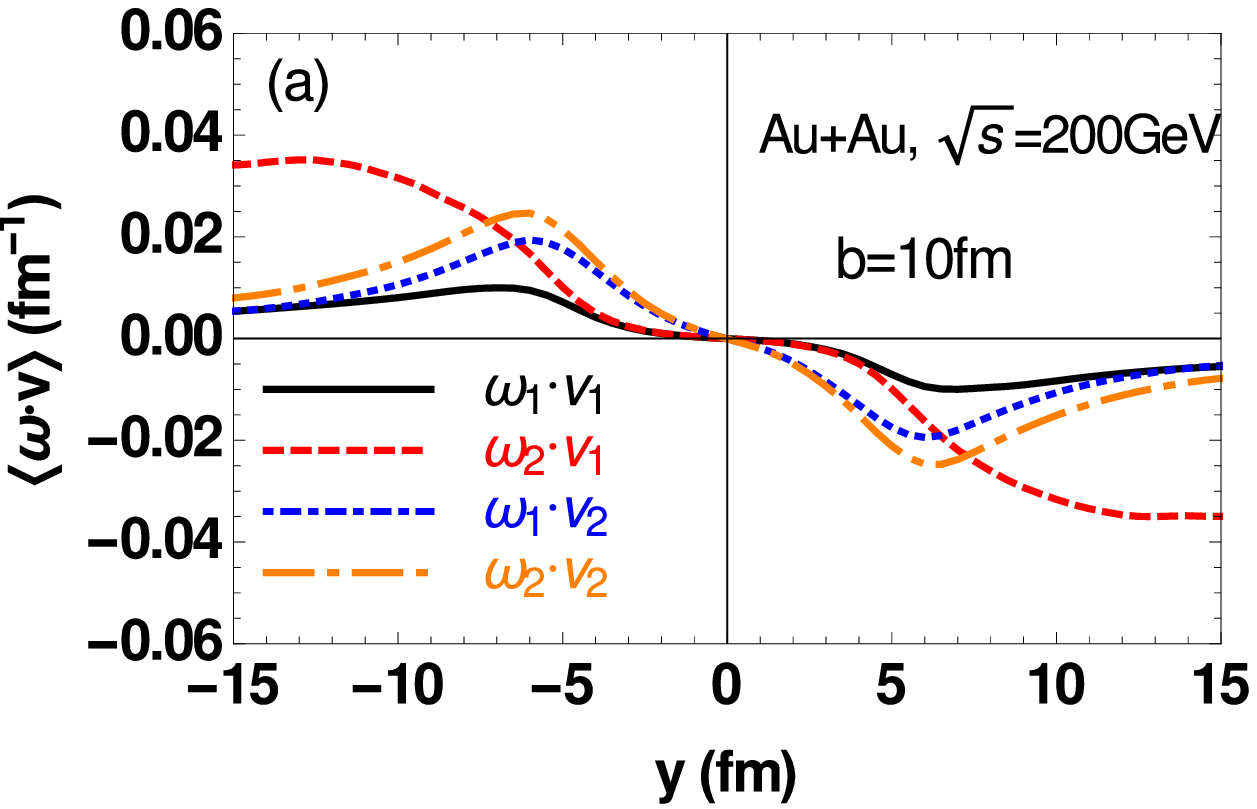}
\includegraphics[width=7cm]{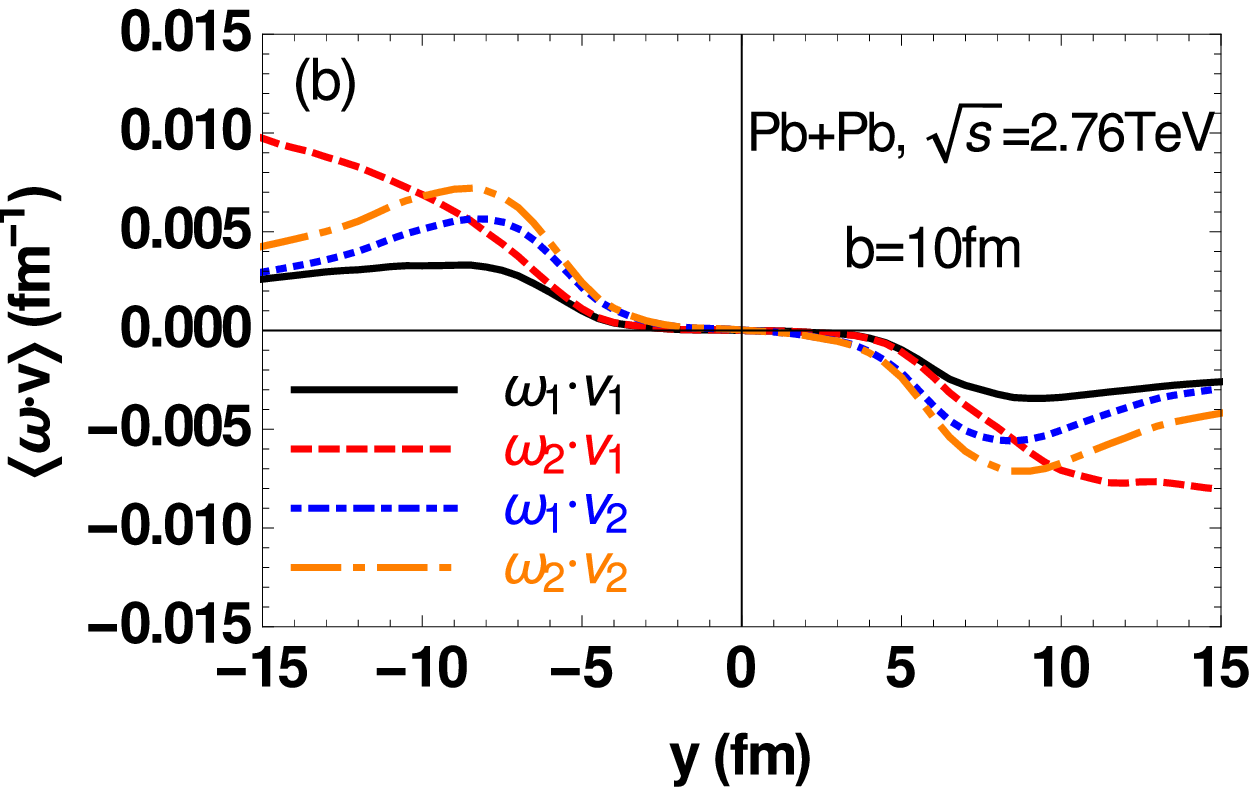}
\caption{The event-averaged helicities $\bv\cdot\vec\o_1$ and $\bv\cdot\vec\o_2$ along $y$ axis. Different curves correspond to different definitions of the vorticity and velocity fields.}
\label{fig-heli-prof}
\end{center}
\end{figure}
\begin{figure}[!htb]
\begin{center}
\includegraphics[width=7cm]{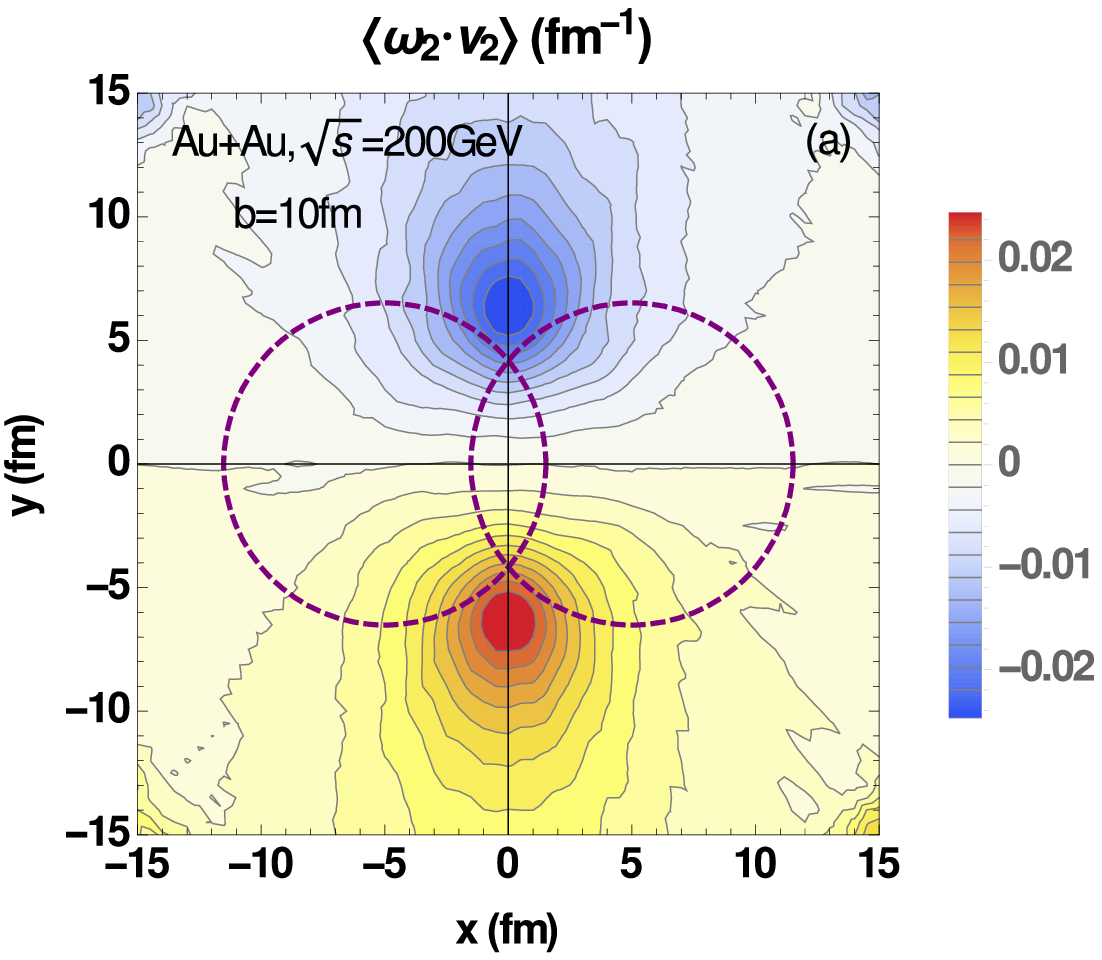}
\includegraphics[width=7cm]{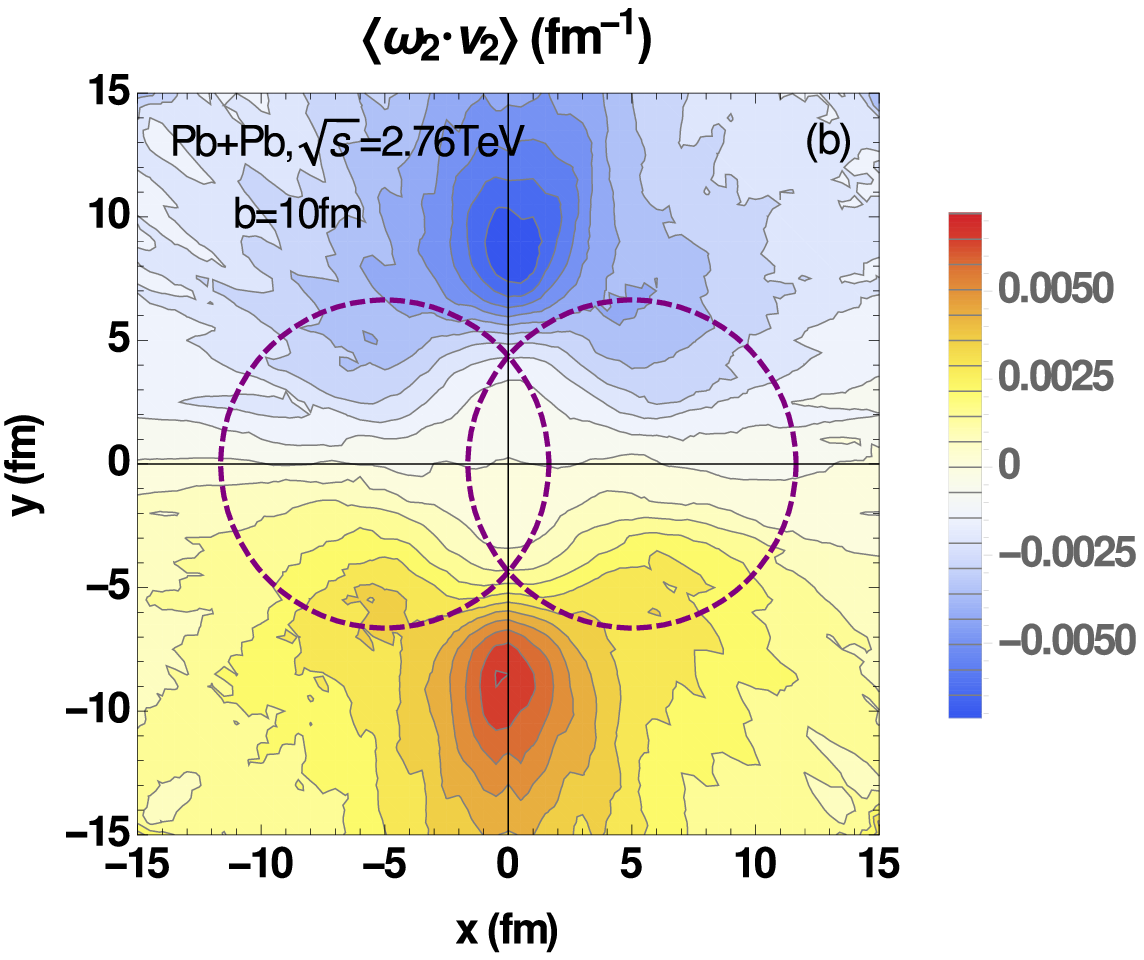}
\caption{The spatial distribution of the event-averaged helicity, $\lan\vec\o_2\cdot\bv_2\ran$, in the transverse plane for RHIC Au + Au collisions at $\sqrt{s}=200$ GeV (panel (a)) and LHC Pb + Pb collisions at $\sqrt{s}=2.76$ TeV (panel (b)).}
\label{fig-heli-spa}
\end{center}
\end{figure}
\begin{figure}[!htb]
\begin{center}
\includegraphics[width=7cm]{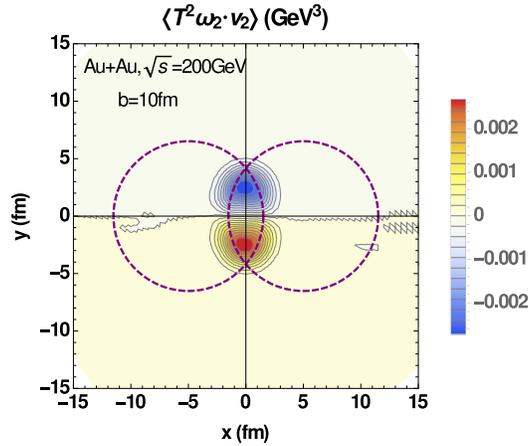}
\caption{The spatial distribution of the event-averaged helicity, $\lan T^2\vec\o_2\cdot\bv_2\ran$, in the transverse plane for RHIC Au + Au collisions at $\sqrt{s}=200$ GeV.}
\label{fig-heli-spa2}
\end{center}
\end{figure}

We in this subsection present the spatial distribution of different kinds of helicity field. In \fig{fig-heli-prof}, we show $\lan\bv\cdot\vec\o_1\ran$ and $\lan\bv\cdot\vec\o_2\ran$ along the $y$ axis for RHIC Au + Au collisions at $\sqrt{s}=200$ GeV and LHC Pb + Pb collisions at $\sqrt{s}=2.76$ TeV. Clearly, the event-averaged helicity is negative for $y>0$ and positive for $y<0$. This becomes more transparent in \fig {fig-heli-spa} where we show the spatial distribution of $\lan\bv\cdot\vec\o_1\ran$ and $\lan\bv\cdot\vec\o_2\ran$ in the transverse plane. Clearly, the reaction plane separates the region with positive event-averaged helicity from the region with negative event-averaged helicity. Similar helicity separation is also observed in low energy heavy-ion collisions~\cite{Baznat:2013zx}. In \fig {fig-heli-spa2}, the $T^2$-weighted helicity $\lan T^2\bv_2\cdot\vec\o_2\ran$ for RHIC Au + Au collisions at $\sqrt{s}=200$ GeV is presented; its physical meaning is given in \sect{relat}. Comparing to $\lan\
{\vec v}\cdot\vec\o_1\ran$ and $\lan\bv\cdot\vec\o_2\ran$, $\lan T^2\bv_2\cdot\vec\o_2\ran$ is much more confined in the overlapping region owing to the fact that $T(\bx_\perp)$ is concentrated in the overlapping region.

The underling mechanism of the helicity separation is simply due to the fact that $\lan v_y\ran$ as a function of $y$ changes its sign from region with $y>0$ to the region with $y<0$ while $\lan \o_y\ran$ does not change the sign. This is similar with the electromagnetic helicity $\bE\cdot\bB$ in heavy-ion collisions where $E_y$ changes its sign from the region below and above the reaction plane but $B_y$ does not~\cite{Deng:2012pc}. The flow helicity separation may have interesting experimental implication, for example, it may generate chiral charges separation via the CVE~\cite{Baznat:2013zx}.

\subsection {Histogram of $\j_\o-\j_2$}\label{subhis}
\begin{figure}[!htb]
\begin{center}
\includegraphics[width=6cm]{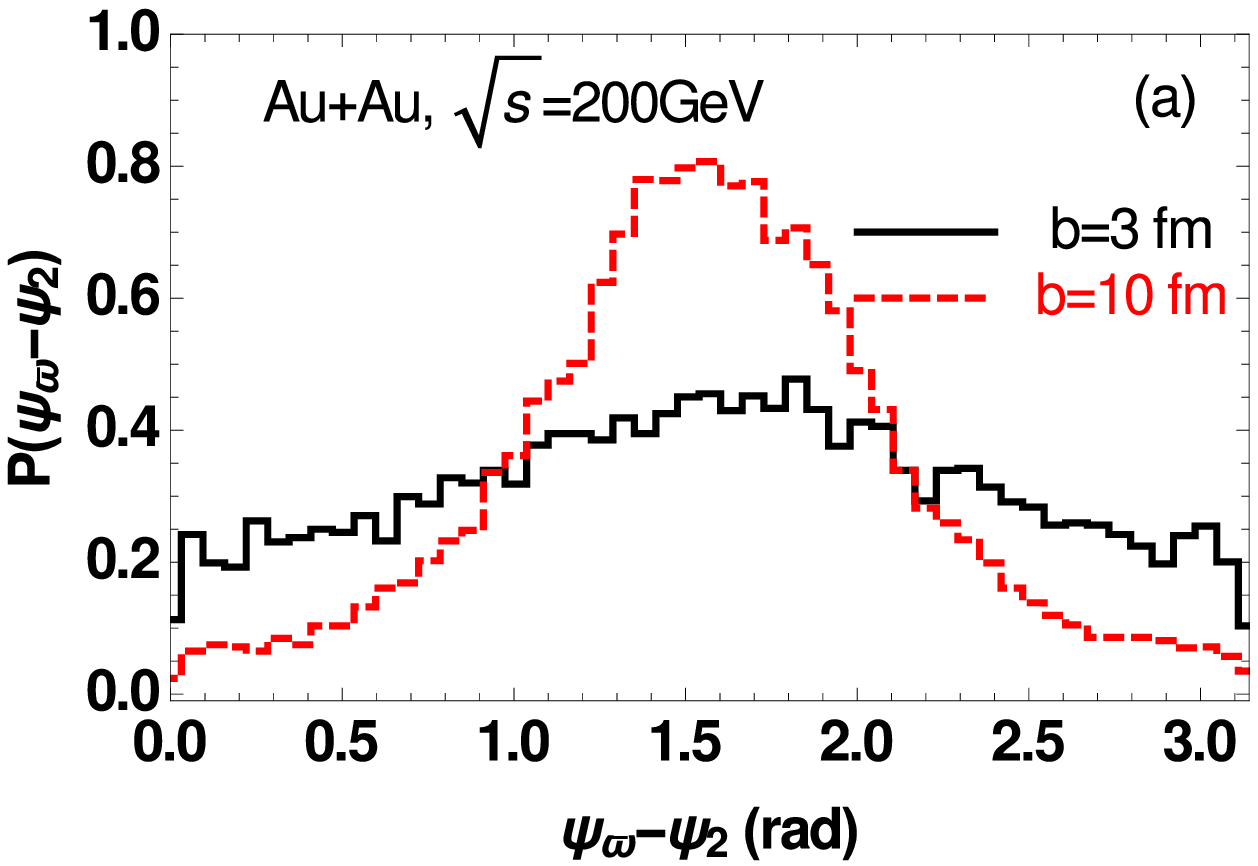}
\includegraphics[width=6cm]{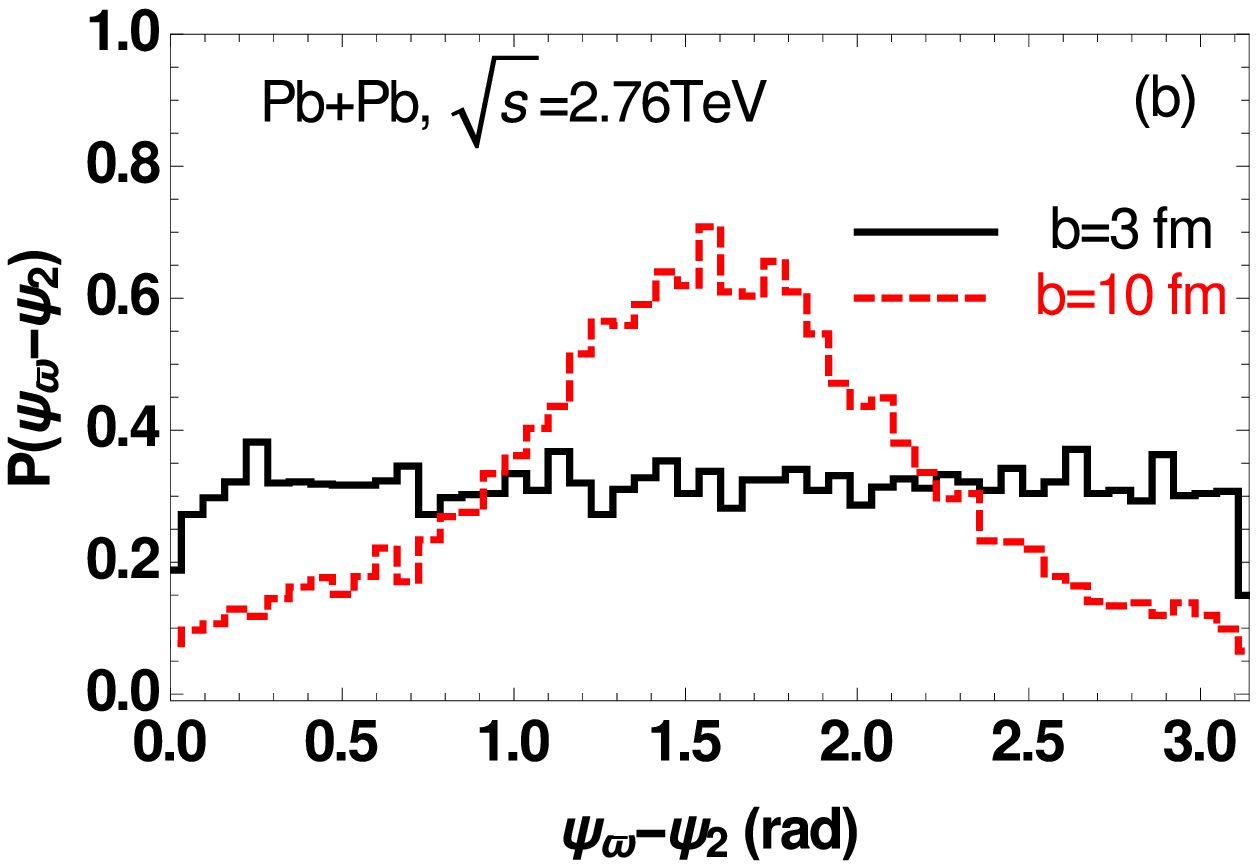}
\caption{The histograms of $\j_\o-\j_2$ at impact parameters $b= 3$ and $10$ fm for Au + Au collisions at RHIC energy (panel (a)) and Pb + Pb collisions at LHC energy (panel (b)). Here $\j_\o$ is the azimuthal direction of the space-averaged vorticity $\bar{\vec\o}_2$ (at $\t_0$ and $\w=0$) of $\bv_2$ and $\j_2$ is the second harmonic participant plane.}
\label{fig-hist}
\end{center}
\end{figure}
As already mentioned in the Introduction, on the event-by-event basis the
vorticity field fluctuates not only in its magnitude but also in its
azimuthal direction. The direction of $\vec\o$ is important in experiments
as the vorticity-driven effects will inherit this information and reflect it in the final obervables. Thus we in this and next subsection will study the event-by-event fluctuation of the azimuthal direction of $\vec\o$ with respect to the matter geometry (more specifically, the participant plane) in detail. Our study will be parallel to the analogous study for electomagnetic fields in Refs.~\cite{Bloczynski:2012en,Bloczynski:2013mca,Deng:2014uja}. For this purpose, we first determine the participant-plane angle (more precisely, the second harmonic component of the participants) $\j_2$ which is known to be firmly correlated to the event plane angle or reaction plane angle. We use the following formula to define $\j_2$ and corresponding eccentricity $\e_2$: $\e_2 e^{in\j_2}=-{\int d^2\bx_\perp\r(\bx_\perp) x_\perp^2 e^{i2\f}/(\int d^2\bx_\perp\r(\bx_\perp) x_\perp^2})$ where $\r(\bx_\perp)$ is the participant density projected onto the transverse plane.

In \fig{fig-hist} we present the histogram of $\j_\o-\j_2$ modulo by $\p$ over $10^5$ events for two different centrality bins, $b=3$ fm and $b=10$ fm, for both RHIC Au + Au collisions at $\sqrt{s}=200$ GeV and LHC Pb + Pb collisions at $\sqrt{s}=2.76$ TeV, where $\j_\o$ is the azimuthal direction of the space-averaged vorticity, $\bar{\vec\o}_2$ based on $\bv_2$ (calculations based on other definitions of the vorticity and velocity show very similar results). The histograms have approximate Gaussian shapes centered at $\j_\o-\j_2=\p/2$ with the corresponding variance widths very large for $b=3$ fm and relatively small at $b=10$ fm. This shows that for central collisions the azimuthal direction of the vorticity suffers from strong event-by-event fluctuation which efficiently kills the correlation between $\j_\o$ and $\j_2$; while for noncentral collisions there is indeed a significant correlations between the two although suppressed by the fluctuation as well. We now turn to more quantitative measure of the
correlation between $\j_\o$ and $\j_2$.

\subsection {Azimuthal correlation between vorticity and participant plane}\label{subazi}
To reveal the azimuthal correlation between the vorticity and the participant plane more quantitatively, we define the following two correlations,
\begin{eqnarray}
\label{R1}
R_1&=&\lan\cos[2(\j_\o-\j_2)]\ran,\\
\label{R2}
R_2&=&\frac{1}{\lan\bar{\vec\o}^2\ran}\lan\bar{\vec\o}^2\cos[2(\j_\o-\j_2)]\ran,
\end{eqnarray}
where $\lan\cdots\ran$ denotes the event average. Similar quantities were used to study the azimuthal correlations between the magnetic field and the participant plane, see Ref.~\cite{Bloczynski:2012en,Bloczynski:2013mca}.
If there is no correlation between the magnitude of the voriticity and its azimuthal direction, $R_2$ should be reduced to $R_1$.

Before showing the numerical results for $R_1$ and $R_2$, we discuss first the physical significance of them. We take the chiral vortical effect (CVE) as an example; other vorticity induced effects can be similarly analyzed. The CVE can induce a baryon number separation along the direction of the voriticity which can be measured through the baryon-number-dependent two-particle correlation,
\begin{eqnarray}
\g_{\a\b}=\lan\cos(\f_\a+\f_\b-2\j_2)\ran,
\end{eqnarray}
where $\a$ (and $\b$) labels the baryon number of the measured particle, i.e., whether the measured particle is a baryon or anti-baryon, and $\f_\a$ is the corresponding azimuthal angle. 
The CVE can induce a special term into the two-particle distribution function of the measured hadrons,
\begin{eqnarray}
f^{\rm CVE}_{\a\b}\propto\vec\o^2\cos(\f_\a-\j_\o)\cos(\f_\b-\j_\o).
\end{eqnarray}
This in turn translates into the following form,
\begin{eqnarray}
f^{\rm CVE}_{\a\b}&\propto&\frac{\vec\o^2}{2}\cos(\f_\a-\f_\b)+\frac{\vec\o^2}{2}\cos[2(\j_\o-\j_2)]\cos(\f_\a+\f_\b-2\j_2)\non
&&-\frac{\vec\o^2}{2}\sin[2(\j_\o-\j_2)]\sin(\f_\a+\f_\b-2\j_2),
\end{eqnarray}
from which we can extract the correlation $\g_{\a\b}$ as
\begin{eqnarray}
\g_{\a\b}\propto\lan\vec\o^2\cos[2(\j_\o-\j_2)]\ran.
\end{eqnarray}
So if the vorticity is perfectly perpendicular to the participant plane, we would have that $\g_{\a\b}$ is proportional to $\vec\o^2$. However, as we have seen from the preceding subsection, this is not the case; the event-by-event azimuthal fluctuation of $\vec\o$ will provide a suppression factor given by $R_2$.
\begin{figure}[!htb]
\begin{center}
\includegraphics[width=6cm]{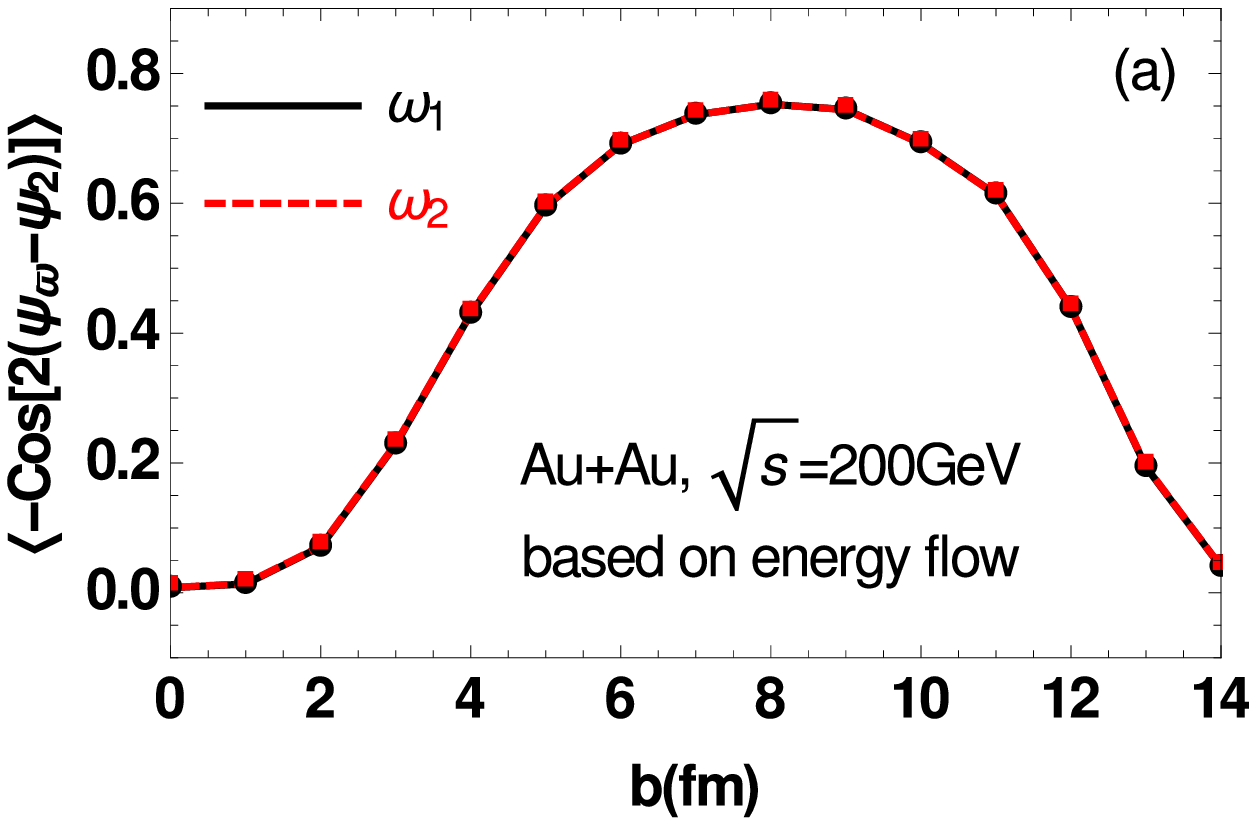}
\includegraphics[width=6cm]{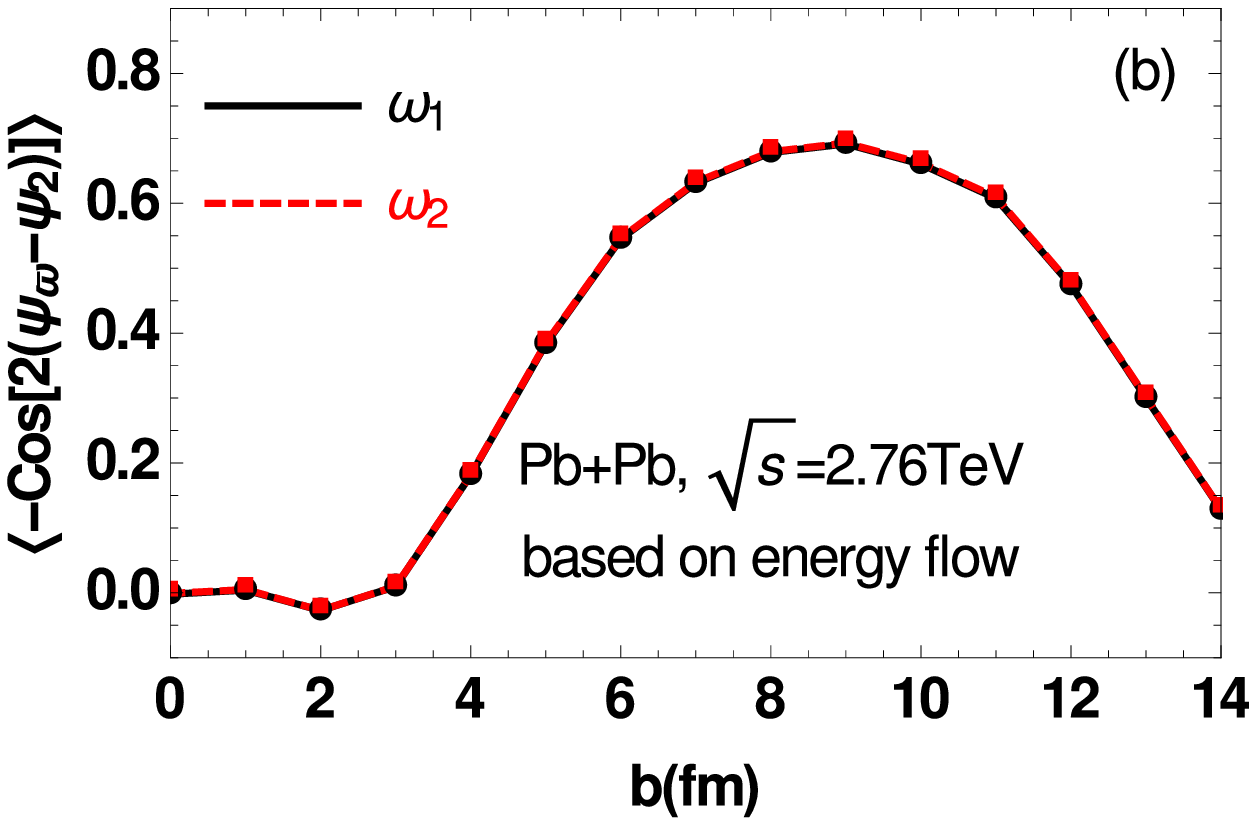}
\caption{The correlation $\lan \cos[2(\j_\o-\j_2)]\ran$ as a function of impact parameter for both RHIC Au + Au collisions at $200$ GeV (panel (a)) and LHC Pb + Pb collisions at $2.76$ TeV (panel(b)). The vorticity field is calculated based on the energy flow velocity $\bv_2$.}
\label{fig-cos-200}
\end{center}
\end{figure}

\begin{figure}[!htb]
\begin{center}
\includegraphics[width=6cm]{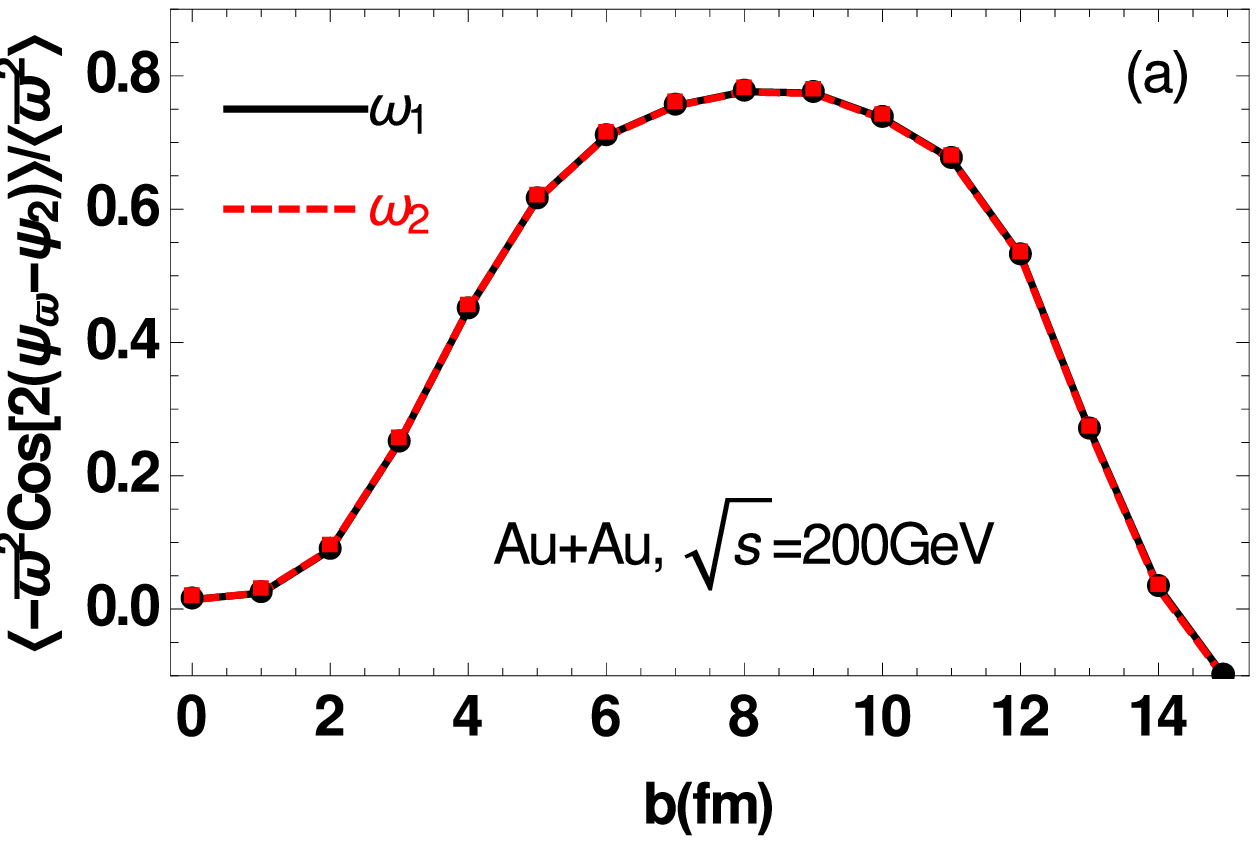}
\includegraphics[width=6cm]{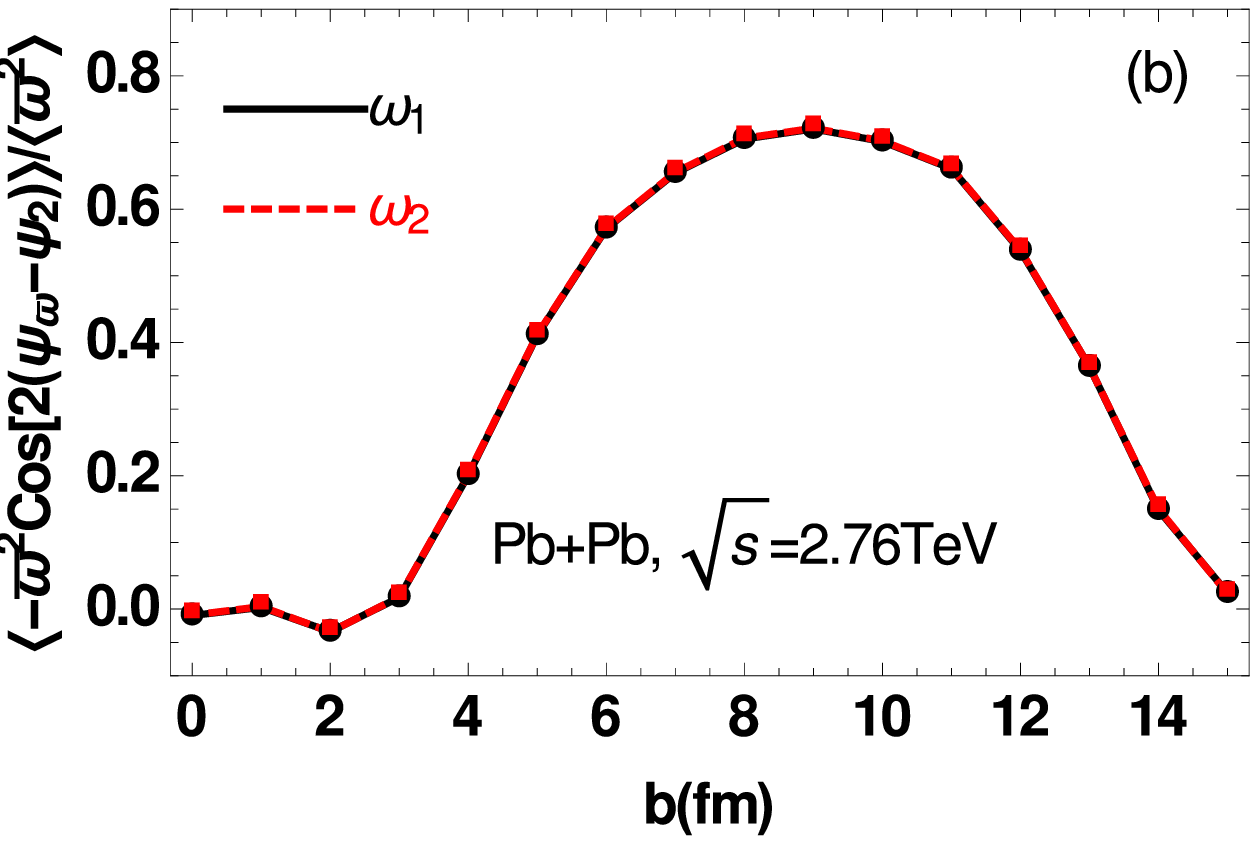}
\caption{The $\vec\o^2$-weighted correlation $\lan \vec\o^2\cos[2(\j_\o-\j_2)]\ran/\lan\vec\o^2\ran$ as a function of impact parameter for both RHIC Au + Au collisions at $200$ GeV (panel (a)) and LHC Pb + Pb collisions at $2.76$ TeV (panel (b)). The vorticity field is calculated based on the energy flow velocity $\bv_2$.}
\label{fig-cos-2.76}
\end{center}
\end{figure}
The correlations $R_1$ and $R_2$ for both RHIC Au + Au collisions and LHC Pb + Pb collisions are presented in \fig{fig-cos-200} and \fig{fig-cos-2.76}. The vorticity fields are calculated based on velocity $\bv_2$; but the results based on velocity $\bv_1$ are qualitatively the same. Evidently, the correlation between $\j_\o$ and $\j_2$ are suppressed comparing to the ideal case without fluctuation, i.e., $\j_\o-\j_2=\p/2$. Both $R_1$ and $R_2$ are significantly suppressed in the most central and most peripheral cases (indicating no strong correlations) and are maximized around $b \sim7 - 10$ fm with peak magnitudes $\sim0.8$ for RHIC and $\sim0.7$ for LHC. Furthermore, we find that practically $R_1\simeq R_2$ for both RHIC and LHC cases suggesting no noticeable correlation between the magnitude and azimuthal direction of $\vec\o$. We note that all these features are very similar with that observed for magnetic field~\cite{Bloczynski:2012en,Bloczynski:2013mca}.

\section {On the time evolution of the vorticity}\label{sectime}
So far, we considered only the vorticity at the fixed proper-time $\t_0$. In this section we turn to discuss the time evolution of the vorticity in the QGP by employing a hydrodynamic analysis. We will not perform full viscous hydrodynamic simulations; instead, our discussion will be based mainly on analytical estimation. We will restrict ourselves to the core domain of the overlapping region where the flow velocity $\bv$ is small and we can use nonrelativistic vorticity ${\bm \o}_1={\bm \nabla}\times{\bm v}$ to proceed with our analysis. To simplify the notation, we will denote $\bm\o_1$ by $\bm\o$ in this section.

The time evolution of the vorticity is goveined by the following vorticity equation~\cite{Landaufluid},
\begin{eqnarray}
\label{eqtime}
\frac{\pt\bm\o}{\pt t}={\vec\nabla}\times(\bv\times\vec\o)+\n\nabla^2\vec\o,
\end{eqnarray}
where $\n=\w/(\ve+P)=T^{-1}(\w/s)$ is the kinematic shear viscosity ($\w$ is the dynamic shear viscosity and $s$ is the entropy density). The first term on the right-hand side is the convection term while the second term represents the diffusion of $\vec\o$ due to shear viscosity. The ratio of these two terms is characterized by the dimensionless Reynolds number,
\begin{eqnarray}
{\rm Re}=UL/\n,
\end{eqnarray}
where $U$ is the characteristic velocity of the flow and $L$ is the characteristic length scale of the fluid. Although \eq{eqtime} is hard to solve in general, we can get important insight via analyzing two limiting cases with ${\rm Re}\ll1$ and ${\rm Re}\gg1$~\footnote{If we assume $U\sim 0.1-1$, $L\sim 5$ fm, $T\sim 300$ MeV, and $\w/s\sim 1/(4\p)$ for a QGP produced at RHIC, $\rm Re\sim 10-100$; at LHC $\rm Re$ would be even larger. Thus it is practically more reasonable to assume $\rm Re\gg 1$.}.

If ${\rm Re}\ll 1$, the convection term can be neglected, and \eq{eqtime} becomes
\begin{eqnarray}
\label{eqdiff}
\frac{\pt\bm\o}{\pt t}=\n\nabla^2\vec\o.
\end{eqnarray}
This is a diffusion equation whose solution is easily obtained by Fourier transformation (suppose that $\n$ is a constant),
\begin{eqnarray}
\vec\o(t,\bk)=\vec\o(0,\bk)e^{-\n\bk^2 t},
\end{eqnarray}
where $\vec\o(t,\bk)$ is the Fourier mode of $\vec\o(t,\bx)$ of wave-number $\bk$. Therefore, the vorticity will decay exponentially with higher wave-number modes decaying faster. More specifically, for illustration purpose, let us consider the initial vorticity distribution in the transverse plane to be a Gaussian,
\begin{eqnarray}
\vec\o(0,\bx)=\vec\o_0 e^{-\bx_\perp^2/\s_r^2},
\end{eqnarray}
with $\s_r$ a width parameter. The solution to \eq{eqdiff} is then
\begin{eqnarray}
\label{otime1}
\vec\o(t,\bx)&=&\int d^3\by\int\frac{d^3\bk}{(2\p)^3} e^{-\n\bk^2 t}\vec\o(0,\by)e^{i\bk\cdot(\bx-\by)}\non
&=&\vec\o_0\frac{\s_r^2}{\s_r^2+4\n t}\exp{\lb-\frac{\bx_\perp^2}{\s_r^2+4\n t}\rb}.
\end{eqnarray}
Thus, for Gaussian initial profile, the vorticity is nearly invariant for $t\ll t_\o=\s_r^2/(4\n)$ and decays exponentially when $t>t_\o$. As we know that $\w/s$ for QGP is quite small, $t_\o$ may be quite long. This is very similar with the previous analysis for the time evolution of the magnetic field in QGP with a large electric conductivity but a small magnetic Reynolds number~\cite{Tuchin:2010vs,Tuchin:2013ie,Tuchin:2014iua}.

If ${\rm Re}\gg 1$, the diffusion term can be neglected, and \eq{eqtime} becomes
\begin{eqnarray}
\label{eqconv}
\frac{\pt\bm\o}{\pt t}={\vec\nabla}\times(\bv\times\vec\o).
\end{eqnarray}
This is just \eq{equvor1}. As well-known, this equation leads to the remarkable Helmholtz-Kelvin theorem of circulation conservation (see Sec.\ref{hydro}). In this case, the vortex lines are frozen in the fluid and the vorticity will decay due to the expansion of the system. To gain some quantitative estimation of this expansion-driven decay, let us decompose the flow velocity into two parts,
\begin{eqnarray}
\label{velocity_decomp}
\bv=\bv_{e}+\bv_c,
\end{eqnarray}
where the first part represents the expansion which we assume to be irrotational, $\vec\nabla\times\bv_e=\vec0$, and the second part represents the vortical flow, $\bv_c=(1/2)\vec\o\times\bx$. Let us consider a small region around the collision center where the vorticity is along the $y$ direction (after event-average) and can be treated as constant (see \fig{fig-vor-spa-200}). Then $\bv_c$ does not contribute to the right-hand side of \eq{eqconv} and \eq{eqconv} becomes
\begin{eqnarray}
\label{eqconv2}
\frac{\pt\bm\o}{\pt t}={\vec\nabla}\times(\bv_e\times\vec\o).
\end{eqnarray}
To proceed, let us assume a Bjorken expansion along the longitudinal direction and a pressure-gradient driven expansion in the transverse plane. Thus
\begin{eqnarray}
v_e^z=\frac{z}{t}.
\end{eqnarray}
Because the early-time transverse expansion is slow, we adopt a linearized ideal hydrodynamic equation to describe it,
\begin{eqnarray}
\label{linearhyd}
\frac{\pt\bv_{e\perp}}{\pt t}=-\frac{1}{\ve+P}\vec\nabla_\perp P=-c_s^2\vec\nabla_\perp\ln s,
\end{eqnarray}
where $c_s=\sqrt{\pt P/\pt\ve}$ is the sound velocity and $s$ is the entropy density. For simplicity, we choose an initial Gaussian profile for $s$,
\begin{eqnarray}
s(\bx_\perp)=s_0\exp{\lb-\frac{x^2}{2a_x^2}-\frac{y^2}{2a_y^2}\rb},
\end{eqnarray}
with $a_{x,y}$ the widths of the transverse entropy distribution. They roughly express the size of the overlap region. For example, for RHIC Au + Au collisions, $a_x\sim a_y\sim 3$ fm at $b=0$ and $a_x\sim 2$ fm and $a_y\sim 3$ fm at $b=10$ fm. One then solves \eq{linearhyd} with~\cite{Ollitrault:2007du}
\begin{eqnarray}
v_{e}^x&=&\frac{c_s^2}{a^2_x}xt,\non
v_{e}^y&=&\frac{c_s^2}{a^2_y}yt.
\end{eqnarray}
Substituting $\bv_e$ into \eq{eqconv2}, we obtain a linear differential equation for $\vec\o(t,\bx)$ which can be solved analytically and gives
\begin{eqnarray}
\label{otime2}
\o_y(t,\bx)=\frac{t_0}{t}\exp{\ls-\frac{c_s^2}{2a_x^2}(t^2-t_0^2)\rs}\o_y(t_0,\bx_0),
\end{eqnarray}
where $\bx_0$ is related to $\bx$ by
\begin{eqnarray}
x&=&x_0\exp{\ls\frac{c_s^2}{2a_x^2}(t^2-t_0^2)\rs},\non
y&=&y_0\exp{\ls\frac{c_s^2}{2a_y^2}(t^2-t_0^2)\rs},\non
\label{solution_z}
z&=&z_0\frac{t}{t_0}.
\end{eqnarray}
These express that a fluid cell located at $\bx_0$ at time $t_0$ flows to $\bx$ at time $t$. The inverse of the prefactor $\frac{t_0}{t}\exp{\ls-\frac{c_s^2}{2a_x^2}(t^2-t_0^2)\rs}$ in \eq{otime2} represents how much the area encircled by a stream line projected to the $x-z$ plane expands from time $t_0$ to $t$, and thus \eq{otime2} is nothing but just the manifestation of the Helmholtz-Kelvin theorem. Especially, at $\bx_0=\vec0$,
\begin{eqnarray}
\label{otime3}
\o_y(t,\vec0)=\frac{t_0}{t}\exp{\ls-\frac{c_s^2}{2a_x^2}(t^2-t_0^2)\rs}\o_y(t_0,\vec0),
\end{eqnarray}
expresses clearly how the vorticity is diluted by the expansion in $x-z$ plane. Setting $a_x\sim a_y\sim 3$ fm, $t_0\sim 0.5$ fm, and $c_s^2\sim 1/3$ for RHIC Au + Au collisions, we find that for $t\lesssim 7$ fm, $\o_y$ is approximately inversely proportional to $t$.

Before we end this section, some comments are in order.

(1) In the case of ${\rm Re}\gg 1$, the Kelvin-Helmholtz instability may be developed which prevents the persistence of a stable laminar flow with finite vorticity~\cite{Drazin:2002}. The underlying mechanism is the circulation conservation. Consider a laminar flow of velocity $v$ along $z$ direction with a constant flow shear $dv/dx$ in $x$ direction (which leads to a constant vorticity in $y$ direction) as an example. Suppose a disturbance is applied at one moment which slightly displaces one layer of the fluid into a sinusoidal shape in the $x-z$ plane. Then the basic laminar flow will drive this sinusoidal layer to be further distorted in such a way that the vorticity at one waist of the sinusoid with positive slope will be transported to its neighboring waist with negative slope (as required by the circulation conservation). The accumulation of the vorticity will then make the disturbance to grow and an instability forms. Detailed analysis shows that the Kelvin-Helmholtz instability grows as $\exp{(at)}$ with $a\propto kU$ ($k$ is the wave-number of the disturbance and $U$ the relative velocity between two fluid layers). Thus the disturbing modes with larger wave-numbers grow faster. In a real fluid, the viscosity is always nonzero which will tend to damp the Kelvin-Helmholtz instability. As the viscous dissipation is stronger for larger wave-number, the competition between the Kelvin-Helmholtz instability and the viscous dissipation sets up a critical wave-number (Kolmogorov scale) above which the viscous dissipation will overcome the Kelvin-Helmholtz instability and drive the hydrodynamic disturbance into thermal fluctuations. In heavy-ion collisions, the Kelvin-Helmholtz instability was studied in detail in Ref.~\cite{Csernai:2011qq} which shows that for small viscosity and large centrality there indeed appears the Kelvin-Helmholtz instability which can drive the fireball to distort in the rapidity direction and can possibly be tested through very careful analysis of the directed flow $v_1$.

(2) In case where the viscosity can be neglected (so that ${\rm Re}\gg 1$), there are known exact solutions to the hydrodynamic equations with rotation (homogeneous vorticity)~\cite{Csernai:2014hva,Nagy:2009eq,Csorgo:2006ax,Csorgo:2015scx}. It is interesting to notice that the rotating Hubble flow given by $\bv=(\bx+t_0{\bm \o}_0\times\bx/2)/t$ with ${\bm \o}_0$ the initial vorticity~\cite{Nagy:2009eq}, which solves the relativistic Euler equation, gives the time evolution of the vorticity in the form of ${\bm \o}(t)=(t_0/t){\bm \o}_0$. Our solution, \eq{velocity_decomp} to \eq{solution_z} which are obtained by solving relativistic hydrodynamic equations although the vorticity equation is nonrelativistic, can be viewed as a rotating Hubble flow expanding in the longitudinal direction plus a transverse expansion due to thermal pressure (In fact, if we turn off the transverse expansion by setting the sound velocity $c_s=0$, our solution is in exactly the form of a rotating Hubble flow expanding in longitudinal direction). In Refs.~\cite{Csorgo:2006ax,Csorgo:2015scx}, the nonrelativistic ideal hydrodynamics with rotation is solved and in this case the decay of vorticity is again driven by the expansion of the system, $\o(t)\sim (R_0^2/R(t)^2)\o_0$ with $R(t)$ the system size transverse to the vorticity direction, but $R(t)$ has a quite nontrivial time dependence.

(3) Finally, we emphasize again that the above analysis is justified only near the collision center; in a region far from the collision center, there would be significant correction due to relativistic flow and the novel spatial distribution of the vorticity. Thus, a full relativistic hydrodynamic simulation is desirable to reveal the detailed time evolution of the vorticity covering more spatial region. An early trial in this direction can be found in Ref.~\cite{Huang:2011ru} where the time evolution of the longitudinal momentum shear is computed in viscous hydrodynamics. Besides, the transport models like the AMPT model may also be used to reveal the time evolution of the vorticity; recently, such a study was performed by the authors of Ref.~\cite{Jiang:2016woz} in which the spatically averaged vorticity is simulated at different moments.

\section {Summary and discussions}\label{discu}
In summary, we have studied the event-by-event generation of the flow vorticity in relativistic heavy-ion collisions by using the HIJING model. To perform the numerical simulation, we have adopted a Gaussian smearing function (\ref{smear2}) to define the velocity field and based on which we have computed the vorticity field. Two types of velocity fields, namely, the particle flow velocity $\bv_1$ and the energy flow velocity $\bv_2$ are defined and two types of vorticity, namely, the nonrelativistic vorticity $\vec\o_1$ and relativistic vorticity $\vec\o_2$ are simulated based on $\bv_1$ and $\bv_2$. From the simulations, we find the following.\\
(1) In non-central relativistic heavy-ion collisions, a sizable fraction of the angular momentum of the two colliding nuclei are accumulated in the collision region. This fraction of angular momentum is manifested in the form of longitudinal flow shear which results in large local flow vorticity. After suitably averaged over the collision region and then over many events, the vorticity is found to be perpendicular to the reaction plane.\\
(2) The vorticity is generally growing with the centrality when the impact parameter $b\lesssim2R_A$ with $R_A$ the radius of the nucleus; for $b>2R_A$ it drops. \\
(3) Although the total angular momentum of the partonic matter increases with increasing collision energy, the event-averaged vorticity decreases with increasing collision energy.\\
(4) For large collision energy, a corona effect is seen in the spatial distribution of the event-averaged vorticity, namely, the maximum vorticity is located around the boundary of the collision region in both the transverse direction and in the spacetime rapidity direction.\\
(5) The event-averaged helicity density exhibits a clear dipolar distribution along the out-of-reaction-plane direction. \\
(6) Both the magnitude and the azimuthal direction of the vorticity suffer from the event-by-event fluctuation. In particular, such fluctuation blurs the vorticity from being perfectly perpendicular to the reaction plane or participant plane. Quantitatively, the absolute values of correlations $R_1$ and $R_2$ are suppressed by the event-by-event fluctuation from being $1$ to at most $0.8$ for RHIC Au + Au collisions at $\sqrt{s}=200$ GeV and $0.7$ for LHC Pb + Pb collisions at $\sqrt{s}=2.76$ TeV.\\
(7) The time evolution of the vorticity is sensitive to the Reynolds number ${\rm Re}$ or equivalently the shear viscosity of the QGP. If ${\rm Re}\ll1$, the vorticity decays due to viscous diffusion. If ${\rm Re}\gg1$, the vortex lines are effectively frozen in the fluid and the vorticity decays due to the hydrodynamic expansion of the QGP.

The presence of vorticity in heavy-ion collisions can have interesting implications in experiment observables, via, for example, the chiral vortical effect and chiral vortical wave. Our study provides an important step towards quantifying these vorticity-driven effects in relativistic heavy-ion collisions, but there are still many aspects of the vorticity as well as the effects it drives that need to be explored, which will be the future tasks.

{\bf Acknowledgments:}
We are grateful to J. Liao, Y. Jiang, and L.-G. Pang for helpful communications and discussions. W.-T.D is supported by the Independent Innovation Research Foundation of Huazhong University of Science and Technology (Grant No. 2014QN190) and the NSFC with Grant No. 11405066. X.-G.H. is supported by NSFC with Grant No. 11535012 and the One Thousand Young Talents Program of China. Part of the numerical computations has been performed at cluster HYPERION in Huazhong University of Science and Technology.

\vskip0.3cm

\appendix
\section{Vorticity for a fluid with a conserved charge}\label{vort-enthapy}
We have noted in \sect{relat} that for relativistic fluid different vorticities can be defined according to the contexts of application. If the fluid carries a conserved charge, one can define the vorticity tensor as
\begin{eqnarray}
\label{enthapyvor}
{\tilde\O}_{\m\n}=\pt_\m( wu_\n)-\pt_\n(wu_\m),
\end{eqnarray}
where $w=(\ve+P)/n$ is the enthalpy per particle with $n$ being the density of the conserved charge. The circulation in correspondence to ${\tilde\O}_{\m\n}$ is
\begin{eqnarray}
\oint w u_\m dx^\m.
\end{eqnarray}
By using the thermodynamic identities, $d\ve=w dn+nTd(s/n)$ and $dP=ndw-nTd(s/n)$ ($s$ is the entropy density), it is straightforward to recast \eq{releuler} to
\begin{eqnarray}
\frac{d}{d\t}(w u^\m)=\pt^\m w+T\nabla^\m(s/n),
\end{eqnarray}
which can be rewritten as the following form (known as the Carter-Lichnerowicz equation)
\begin{eqnarray}
\label{CLequ}
\tilde{\O}_{\m\n} u^\n=T\nabla_\m(s/n).
\end{eqnarray}
Thus for isentropic flow, i.e., $s/n$ is strictly constant, one obtains the following circulation conservation:~\footnote{The isentropic condition can be relaxed. In fact, a weaker version of circulation conservation can hold following directly the Carter-Lichnerowicz equation~\cite{Katz:1984,Bekenstein:1987}. }
\begin{eqnarray}
\frac{d}{d\t}\oint w u_\m dx^\m=\oint \pt_\m wdx^\m=0.
\end{eqnarray}
Define a (pseudo)vector field
\begin{eqnarray}
{\tilde\O}^\m=\frac{1}{2}\e^{\m\n\r\s}wu_\n \tilde{\O}_{\r\s}=w^2\o_2^\m.
\end{eqnarray}
Its divergence reads (a consequence of the Carter-Lichnerowicz equation),
\begin{eqnarray}
\pt_\m\tilde{\O}^\m=\frac{1}{2}\e^{\m\n\r\s}\tilde{\O}_{\m\n}\tilde{\O}_{\r\s}=-2 \frac{T}{w}\tilde{\O}^\m\nabla_\m \frac{s}{n}.
\end{eqnarray}
Thus for isentropic fluid we have $\pt_\m\tilde{\O}^\m=0$ which implies that $\int d^3\bx {\tilde\O}^0$ is conserved. This is the relativistic version of the helicity conservation for fluid with a conserved charge.

\section {Transformation between Cartesian and proper-time coordinates}\label{append-coo}
In the Cartesian coordinates, $x^\m=(t,\vec x)$, $g_{\m\n}=g^{\m\n}=\diag(1,-1,-1,-1)$. From energy-momentum tensor of ideal fluid (where $\m,\n=0,1,2,3$)
\begin{eqnarray}
\label{append-em}
T^{\m\n}=(\ve+P)u^\m u^\n-Pg^{\m\n},
\end{eqnarray}
where $P$ and $\ve$ are pressure and energy density, and $u^\m=\g(1,\bv)$ is the velocity of energy flow:
\begin{eqnarray}
T^{\m\n}u_\n=\ve u^\m.
\end{eqnarray}
From \eq{append-em}, we obtain
\begin{eqnarray}
\frac{v^a}{1+(v^a)^2}=\frac{T^{0a}}{T^{00}+T^{aa}},
\end{eqnarray}
where $a=1,2,3$ and the repeated indices are not summed. Solving this equation for $v^a$, we get
\begin{eqnarray}
v^a=\frac{1-\sqrt{1-4V^2}}{2V},
\end{eqnarray}
with $V\equiv T^{0a}/(T^{00}+T^{aa})$. In case that $V$ is not large, we have
\begin{eqnarray}
v^a\approx \frac{T^{0a}}{T^{00}+T^{aa}}.
\end{eqnarray}
This expression is used to define $\bv_2$ is \sect{comp}.

In the proper-time coordinates, $\tilde{x}^\m=(\t,x,y,\w)$ with the proper time $\t=\sqrt{t^2-z^2}$ and spacetime rapidity $\w=(1/2)\ln[(t+z)/(t-z)]$ or inversely $t=\t\cosh\w$ and $z=\t\sinh\w$. The corresponding metric is $\tilde{g}_{\m\n}=\diag(1,-1,-1,-\t^2)$ and its inverse is $\tilde{g}^{\m\n}=\diag(1,-1,-1,-1/\t^2)$. Let $A^{\m}(x)$ be a vector written in the Cartesian coordinates and its corresponding counterpart in the proper-time coordinates is $\tilde{A}^{\m}(\tilde{x})$. The transformation between $A^{\m}$ and $\tilde{A}^\m$ is given by
\begin{eqnarray}
A^0(x)&=&\frac{t}{\t}\tilde{A}^\t(\tilde x)+z\tilde{A}^\w(\tilde x),\non
A^x(x)&=&\tilde{A}^x(\tilde x),\non
A^y(x)&=&\tilde{A}^y(\tilde x),\non
A^z(x)&=&\frac{z}{\t}\tilde{A}^\t(\tilde x)+t\tilde{A}^\w(\tilde x).
\end{eqnarray}
Or in a compact form, $A^\m(x)={\L^\m}_\n \tilde{A}^\n[\tilde{x}(x)]$, with the transformation matrix given by
\begin{eqnarray}
\displaystyle ({\L^\m}_\n)=\lb\begin{matrix}\frac{t}{\t}, 0, 0, z\\ 0, 1, 0, 0\\0, 0, 1, 0\\\frac{z}{\t}, 0, 0, t\end{matrix}\rb.
\end{eqnarray}
By using ${\L^\m}_\n$, it is easy to find the relation between $T^{\m\n}(x)$ and $\tilde{T}^{\m\n}(\tilde{x})$:
\begin{eqnarray}
T^{00}&=&\frac{t^2}{\t^2}\tilde{T}^{\t\t}+\frac{2tz}{\t}\tilde{T}^{\t\w}+z^2\tilde{T}^{\w\w},\non
T^{0x}&=&\frac{t}{\t}\tilde{T}^{\t x}+z\tilde{T}^{\w x},\non
T^{0y}&=&\frac{t}{\t}\tilde{T}^{\t y}+z\tilde{T}^{\w y},\non
T^{0z}&=&\frac{tz}{\t^2}\tilde{T}^{\t\t}+\frac{t^2+z^2}{\t}\tilde{T}^{\t\w}+tz\tilde{T}^{\w\w},\non
T^{xx}&=&\tilde{T}^{xx},\non
T^{yy}&=&\tilde{T}^{yy},\non
T^{zz}&=&\frac{z^2}{\t^2}\tilde{T}^{\t\t}+\frac{2tz}{\t}\tilde{T}^{\t\w}+t^2\tilde{T}^{\w\w}.
\end{eqnarray}
These relations are used in our computations. Especially, at zero rapidity, the two coordinate systems coincide with each other which simplifies our computations.

\section {Another method to extract the velocity field}\label{newsmearing}
The numerical result for the velocity field depends on the choice of the smearing function $\F(x,x_i)$. In the main text, we use the Gaussian smearing method. In this Appendix, we discuss another smearing function which generalizes $\d^{(3)}[\bx-\bx_i(t)]$ (which corresponds to zero smearing) to
\begin{eqnarray}
\label{smear1}
\F_\D(x,x_i)=\d_\D^{(3)}(\bx-\bx_i(t)),
\end{eqnarray}
which is defined as follows. If $|x-x_i(t)|<\D x, |y-y_i(t)|<\D y, |z-z_i(t)|<\D z$, then $\d_\D^{(3)}(\bx-\bx_i(t))=1$; otherwise it is zero. In practical simulation, such a smearing can be achieved by discretizing the space into small cells of volume $\D x \D y \D z$ and the velocity at point $\bx$ is set to be the velocity of the cell that $\bx$ belongs to. Such a smearing is widely used in transport models. Recently, the voticity field was computed in Hadron-String Dynamics model~\cite{Teryaev:2015gxa} and A Multi-Phase Transport(AMPT) model~\cite{Jiang:2016woz} by using such a method to define the velocity field.

\begin{figure}
\begin{center}
\includegraphics[width=7cm]{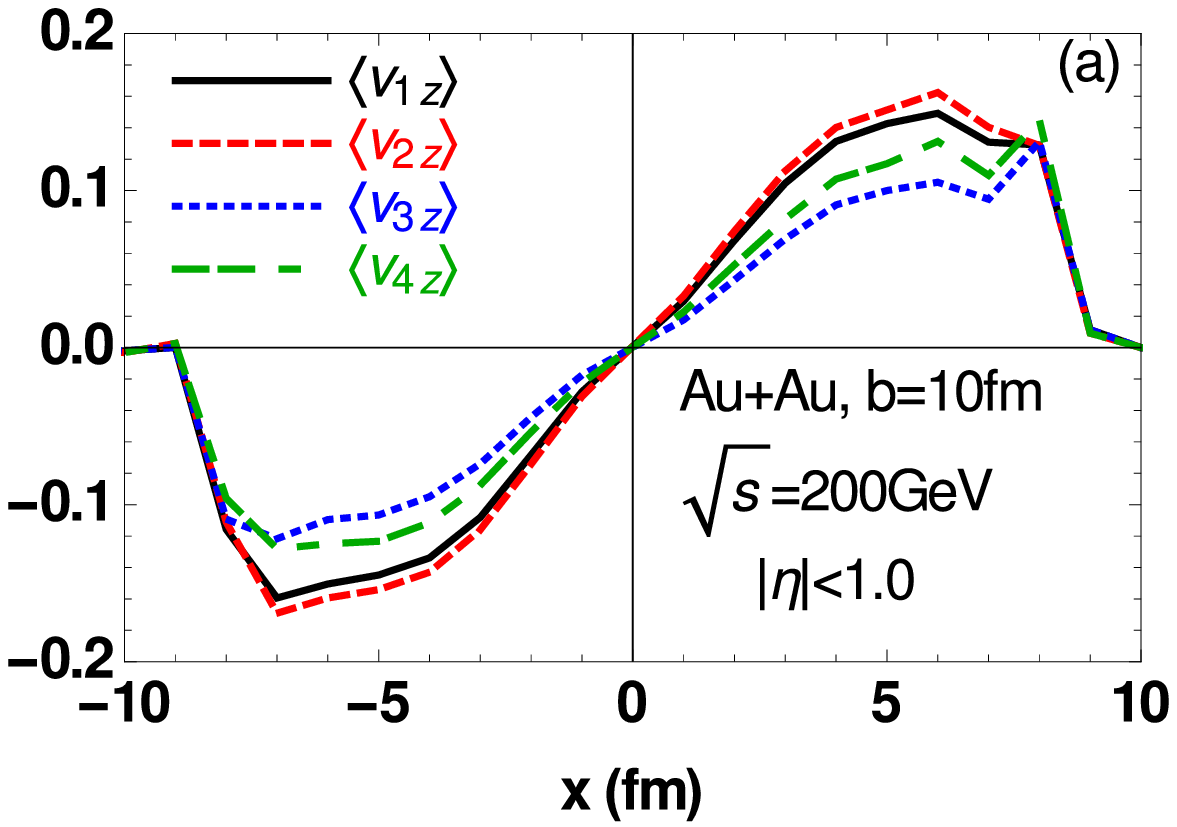}
\includegraphics[width=7cm]{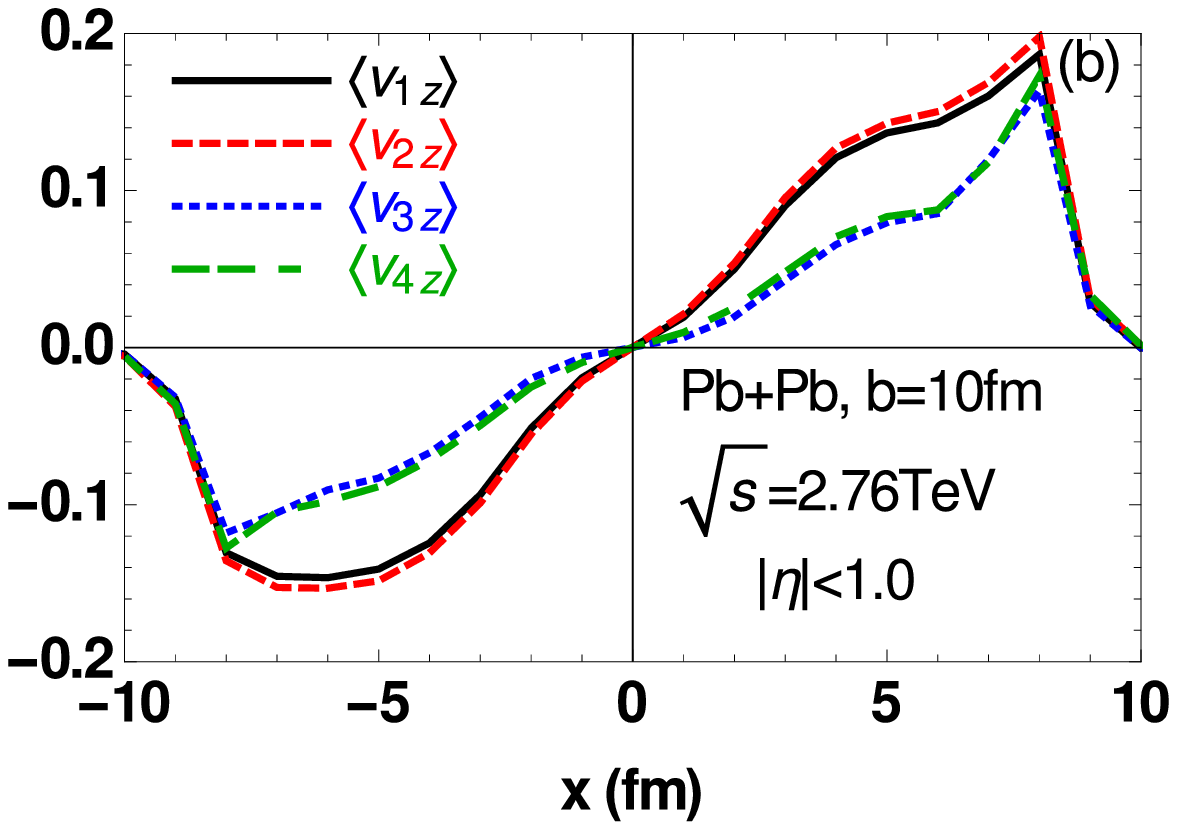}
\caption{The event-averaged longitudinal velocity profile at zero rapidity for RIHC (panel (a)) and LHC (panel (b)). Different curves correspond to different definitions of the event-averaged velocity, see \eq{v-12} $-$ \eq{v-12p2}.}
\label{fig-vz-cell}
\end{center}
\end{figure}

The event-averaged longitudinal velocity profile at $\t=0$ computed by using the above method is shown in \fig{fig-vz-cell} which we run $10^7$ events and choose $\D x=\D y=1$ fm and $\D z=\infty$. We checked that varying $\D x$ and $\D y$ from $0.1$ fm to $2$ fm results no more than $10\%$ variation in velocity. The behavior at small $s$ is similar with the result obtained by using $\F_{\rm G}$. At large $x$, the two smearing methods give different results. Particularly, $\F_\D$ does not lead to finite $\lan v_z\ran$ for $x>R_A+b/2$ where $R_A$ is the radius of the nucleus.


\end{document}